%% file: PTOLEMYTP.tex
\documentclass[12pt]{article}
\usepackage[pdftex]{graphicx}
\usepackage{bibunits}
\usepackage{units}

\usepackage[section]{placeins}
\usepackage{tocbibind}
\usepackage{notoccite}

\usepackage{hyperref}
\usepackage{latexsym}
\usepackage{xspace}
\usepackage{relsize}
\usepackage{subfigure}
\usepackage{authblk}
\input{./localsym}

\begin{document}
\title{\normalsize Development of a Relic Neutrino Detection Experiment at PTOLEMY: \\ Princeton Tritium Observatory for Light, Early-Universe, Massive-Neutrino Yield}
\author[1]{\normalsize S. Betts}
\affil{\small Princeton Plasma Physics Laboratory}
\author[1]{\normalsize W. R. Blanchard}
\author[1]{\normalsize R. H. Carnevale}
\author[2]{\normalsize  C. Chang}
\affil{\small Argonne National Laboratory and University of Chicago}
\author[3]{\normalsize C. Chen}
\affil{\small Department of Physics, Princeton University}
\author[3]{\normalsize S. Chidzik}
\author[1]{\normalsize L. Ciebiera}
\author[4]{\normalsize P. Cloessner}
\affil{\small Savannah River National Laboratory}
\author[5]{\normalsize A. Cocco}
\affil{\small Istituto Nazionale di Fisica Nucleare -- Sezione di Napoli}
\author[1]{\normalsize A. Cohen}
\author[1]{\normalsize J. Dong}
\author[3]{\normalsize R. Klemmer}
\author[3]{\normalsize M. Komor}
\author[1]{\normalsize C. Gentile}
\author[3]{\normalsize B. Harrop}
\author[1]{\normalsize A. Hopkins}
\author[3]{\normalsize N. Jarosik}
\author[5]{\normalsize G. Mangano}
\author[6]{\normalsize M. Messina}
\affil{\small Department of Physics, Columbia University}
\author[3]{\normalsize B. Osherson}
\author[1]{\normalsize Y. Raitses}
\author[3]{\normalsize W. Sands}
\author[1]{\normalsize M. Schaefer}
\author[1]{\normalsize J. Taylor}
\author[3]{\normalsize C. G. Tully}
\author[1]{\normalsize R. Woolley}
\author[1]{\normalsize A. Zwicker}

\date{}

\maketitle
\begin{figure}[h!]
\begin{center}
\includegraphics[width=0.6\textwidth]{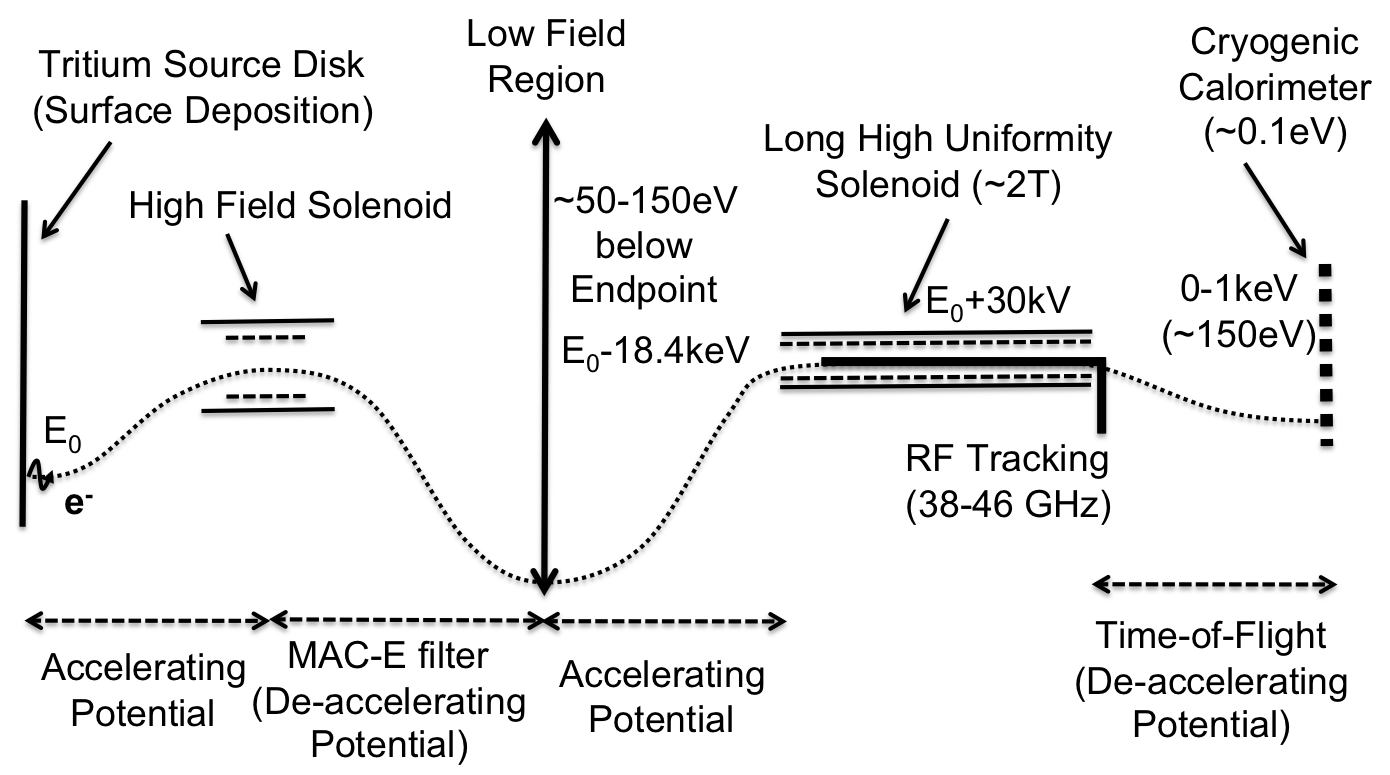}
\caption{The PTOLEMY conceptual design starts with a large area surface-deposition tritium source, accelerates into a MAC-E filter with $10^{-3}$--$10^{-4}$ cut-off precision, accelerates electrons above the endpoint and down to 50--150eV below the endpoint into a long, uniform field solenoid where the RF signal from the cyclotron motion of individual electrons in a 2T magnetic field provide a tracking detector measurement above a minimum transverse momentum, then finally the electron is decelerated into a sub-keV energy range, low magnetic field region, and measured with a high resolution cryogenic calorimeter in time-of-flight coincidence with the RF tracker.
\label{fig:PTOLEMYschematic}}
\end{center}
\end{figure}

\begin{figure}[h!]
\begin{center}
\includegraphics[width=0.6\textwidth]{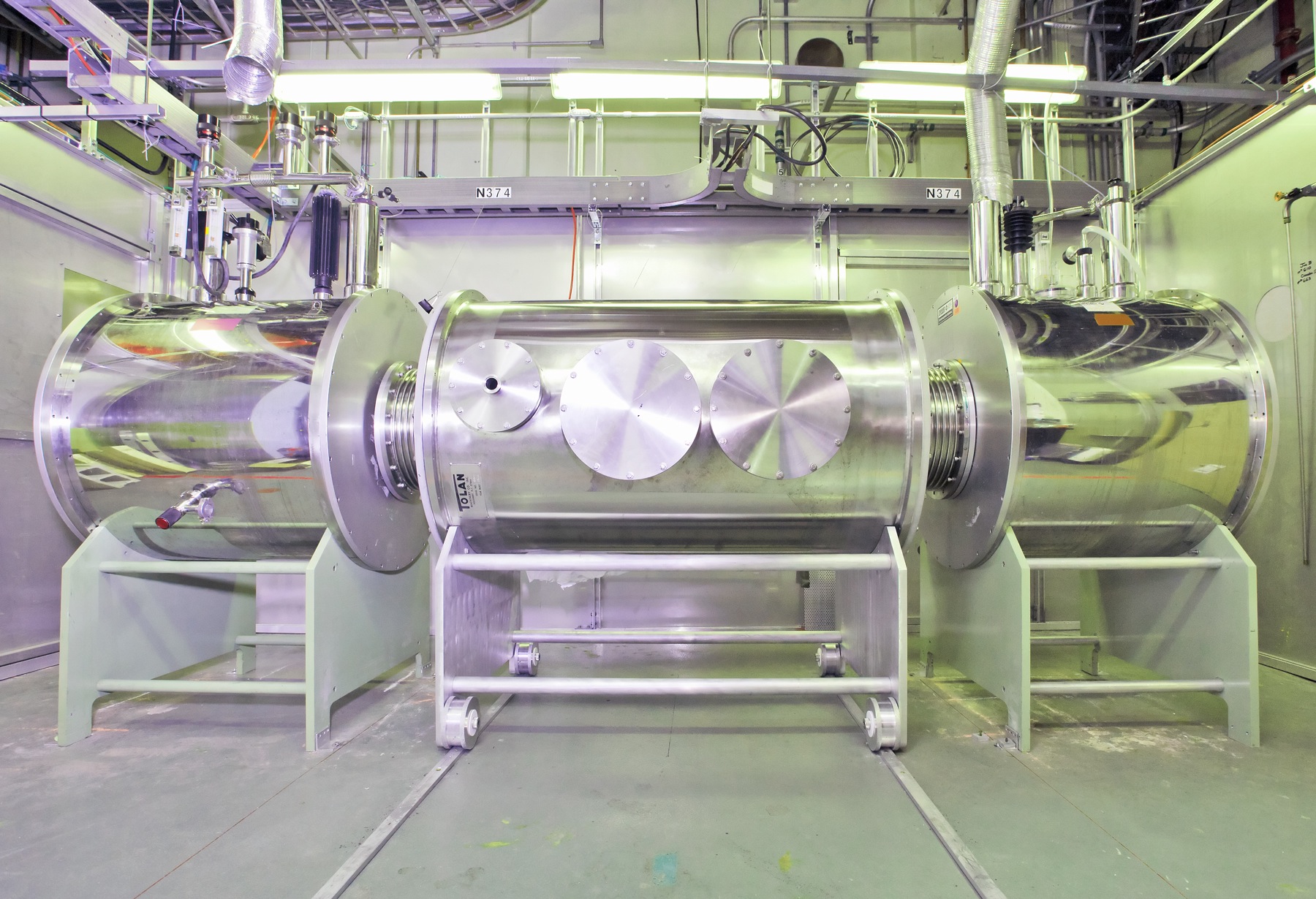}
\caption{The small-scale PTOLEMY prototype installed at the Princeton Plasma Physics Laboratory (February 2013).  Two horizontal bore NMR magnets are positioned on either side of a MAC-E filter vacuum tank.  The tritium target
plate is placed in the left magnet in a 3.35T field, and the RF tracking system is placed in a high uniformity 1.9T field
in the bore of the right magnet with a windowless APD detector and in-vacuum readout electronics.
\label{fig:PTOLEMYphoto}}
\end{center}
\end{figure}

 
\centerline{\bf Project Summary}
The direct detection of relic neutrinos from the Big Bang was proposed in a paper by Steven Weinberg in 1962 [{\it Phys. Rev.} 128:3 (1962) 1457].  The signal for relic neutrino capture on tritium is the observation of electron kinetic energies emitted from a tritium target that are above the $\beta$-decay endpoint.  The requirements on the experimental energy resolution for relic neutrino identification are constrained by the thermal model for neutrino decoupling in the early universe that predicts a present-day average neutrino kinetic energy of $1.7\times 10^{-4}$eV, neutrino mass mixing parameters that indicate mass eigenstates at least as massive as 0.05eV, and cosmological input from WMAP+SPT, and other sources, on the sum of the masses
of the light neutrino species in thermal equilibrium in the early universe to be constrained to less than approximately 0.3eV.
The parameters for a relic neutrino experiment require 100 grams of weakly-bound atomic tritium, sub-eV energy resolution commensurate with the most massive neutrinos with electron-flavor content, and below microHertz of background rate in a narrow energy window above the tritium endpoint.
The PTOLEMY experiment (Princeton Tritium Observatory for Light, Early-Universe, Massive-Neutrino Yield) aims to achieve these goals through a combination of a large area surface-deposition tritium target, MAC-E filter methods, cryogenic calorimetry, and RF tracking and time-of-flight systems.  A schematic of the PTOLEMY concept is shown in figure~\ref{fig:PTOLEMYschematic}.
A small-scale prototype is in operation at the Princeton Plasma Physics Laboratory, shown in figure~\ref{fig:PTOLEMYphoto}, with the goal of validating the technologies that would enable the design of a 100 gram PTOLEMY.  With precision calorimetry in the prototype setup, the limitations from quantum mechanical and Doppler broadening of the tritium target for different substrates will be measured, including graphene substrates.  Beyond relic neutrino physics, sterile neutrinos contributing to the dark matter in the universe are allowed by current constraints on partial contributions to the number of active neutrino species in thermal equilibrium in the early universe.
The current PTOLEMY prototype is expected to have unique sensitivity in the search for sterile neutrinos with
electron-flavor content for masses of 0.1--1keV, where less stringent, 10eV, energy resolution is required.
The search for sterile neutrinos with electron-flavor content with the 100g PTOLEMY is expected
to reach the level $|U_{e4}|^2$ of $10^{-4}$--$10^{-6}$, depending on the sterile neutrino mass.

\vspace*{1cm}
\pagenumbering{roman}
\tableofcontents

\newpage
\input{./sections/Introduction/introduction}

\input{./sections/Performance/performance}

\input{./sections/Tritium/tritium}
\input{./sections/MAC-E/mac-e}

\input{./sections/RF/rf}

\input{./sections/Calorimeter/calorimeter}

\input{./sections/TriggerDAQ/triggerdaq}

\input{./sections/TOF/tof}

\input{./sections/MuonVeto/muonveto}

\input{./sections/Calibration/calibration}

\input{./sections/Vacuum/vacuum}

\input{./sections/Cooling/cooling}
\input{./sections/Monitoring/monitoring}
\input{./sections/Support/support}

\input{./sections/Safety/safety}

\input{./sections/ScienceEd/science}
\input{./sections/Prototype/prototype}

\bibliographystyle{ptolemy-bibstyle}
\bibliography{PTOLEMYTP}

\end{document}

%% file: sections/Introduction/introduction.tex
\pagenumbering{arabic}
\vspace{-0.25 truein}
\section{Introduction}

A possible thermal history of the universe was put forth in a paper by
Dicke, Peebles, Roll, and Wilkinson in 1965 [{\it Astro.J.} 142:2 (1965) 419].
Light neutrinos produced in the early universe are predicted to have thermally decoupled
from other forms of matter at approximately 1~second after the Big Bang.  The light neutrinos, 
believed to be stable particles, have cooled to a temperature of 1.9K (1.7$\times$10$^{-4}$eV) in the 
present day and are predicted to have an average number of density of approximately 56/cm$^3$ per 
lepton flavor.  The number density prediction is based on the annihilation rate of $e^+e^-$ into three
flavors of neutrinos through the weak neutral current interaction in the dense, high temperature 
conditions of the early universe.

A method for direct detection of the relic neutrinos with electron flavor was originally proposed 
in a paper by Steven Weinberg in 1962 [{\it Phys. Rev.} 128:3 (1962) 1457].  The technique is based
on the process of neutrino capture on tritium.  Neutrino capture on a tritium target consists of a relic neutrino 
and a tritium nucleus in the initial state, and two particles in the final state, an electron and
a $^3$He nucleus.   If the kinetic energy of the electrons emitted from tritium nuclei are measured
with a precision comparable to the masses of the relic neutrino eigenstates, then the process of tritium 
$\beta$-decay, a three-body final state, can be distinguished from neutrino capture by looking
for electron energies above the $\beta$-decay endpoint with a separation set by twice the neutrino mass, for
the neutrino mass eigenstates with electron flavor.  
An extensive survey of possible target nuclei 
was conducted by Cocco, Mangano, and Messina in 2007 [{\it JCAP} 0706:015, hep-ph/0703075].
The uncertainty on the neutrino capture cross section on tritium is constrained to the sub-percent level,  
and among the possible
nuclei unstable to $\beta$-decay the tritium nuclei is considered optimal based on the product of the
capture cross section and the half-life of 12.3~years uniquely coming from the $\beta$-decay process.
A tritium target mass of 100~grams is predicted to produce approximately 9.51$\pm$0.03~event/year from 
relic neutrino capture with the relic neutrino density modeled as a uniform Fermi-Dirac number density 
throughout space.  The local neutrino density may be enhanced in galactic clusters by factors that range 
from 1-100 depending on the neutrino mass [Ringwald and Wong, 2004].

A tritium observatory for light neutrinos from the early universe has to achieve a target mass exposure
of at least 100~gram-years to measure the nominal number density expected from Big Bang cosmology.  Furthermore, the energy smearing of the electrons produced from the capture process must be below the 
comparable neutrino mass sensitivity needed to separate out the tritium $\beta$-decay background.
The thermal and quantum mechanical properties of tritium atoms serving as target nuclei for the
capture of relic neutrinos are potentially leading sources of electron energy smearing.
The binding energy of the tritium atom and the possible $^3$He bound states
introduce a shift and spread in the kinetic energy of an endpoint electron.  The energy spread from
averaging over the multiplicity of available quantum mechanical transitions is minimized by weakly binding 
the tritium atom to a nearly inert chemical substrate with a binding energy comparable to the required energy resolution on the endpoint energy.  This is in contrast to tritium atoms in covalent bonds such as in molecular
T$^2$.  The {\it graphane} structure, single atomic-layer graphene with weakly-bound hydrogen (or tritium)
in the $sp$-3 configuration with the carbon lattice, is one such
substrate (http://www.sciencemag.org/content/323/5914/610.abstract).  The binding energy of the tritium
is sub-eV as compared to over 4eV in T$^2$.  This is demonstrated by the annealing of hydrogen
off of graphene at 700K with a measured desorption energy of 19.6kJ/mol 
[Dillon and others, {\it Nature}, Vol 386 (1997) p. 377].
The range of final states with $^3$He has a similar reduction in the scale of the
shifts and smearing on the electron endpoint energy.  The substrate temperature from studies of T$^2$
is roughly a 0.2eV smearing at 20--30K, and therefore substrate temperatures an order of magnitude lower
bring this source of energy smearing to a sub-dominant level.  The thickness of the tritium target is limited by
the scattering probability of the outgoing electron with other tritium or carbon atoms.  In fully tritrated graphane, the tritium attaches on both sides of graphene with a spacing of 2.42$\AA$ or approximately $3 \times 10^{15}$ tritium 
nuclei/cm$^2$.  The cross section of $3 \times 10^{-18}$~cm$^2$ for 18.6keV electrons to scatter with tritium atoms therefore limits the thickness of a tritium target to $\sim$10--100 atomic tritium layers to keep the scattering probability low.  With a graphene substrate, the $Z^2$ dependence of the source scattering limits the number of stacked graphene
layers down to 2--3.
This corresponds to roughly 1$\mu$gram of tritium per cm$^2$.  The total effective area of the tritium target for 100grams is 10$^8$cm$^2$.  Therefore, for a nominal planar target surface of 100m$^2$, a non-planar topology increasing the effective source area by a factor of 100 is needed to achieve the required target mass.

The low expected signal rate of approximately 0.3 microHertz of neutrino capture events places tight constraints on the allowed background rate in the experiment.  The non-electron backgrounds interacting directly with the calorimeter are numerous in origin and would overwhelm the signal without the incorporation of tight electron identification criteria.  The primary method of background rejection is from a narrow energy window at the endpoint limited in width by the energy resolution of the calorimeter.  
Further rejection of background with the calorimeter data alone is limited to pulse shape information and regional hit multiplicity.
The addition of RF tracker techniques developed for the Project 8 neutrino mass measurement [Monreal and Formaggio (2009) arXiv/904.2860, doi:10.1103/PhysRevD.80.051301] provides unique capabilities for electron identification through the detection of electromagnetic radiation from the cyclotron motion at a known frequency for semi-relativistic electrons at the tritium endpoint.  The time-of-flight between the RF tracker and the calorimeter provides a powerful coincidence and selection criteria for endpoint electrons by combining the semi-relativistic electron radiation signature, the tracking trajectory through the magnetic geometry between the tritium target and the calorimeter, and the narrow energy resolution of the calorimeter.  
These constraints suppress non-electron backgrounds down to the level of random coincidence between tracker and calorimeter signals.
Different sources of electron background range from tritium $\beta$-decay and scattering in the target, 
$\beta$-decay and EC electrons from unstable isotope contaminants in the target substrate and the opening of the 
tracker region, pair production from $\gamma$ sources on these surfaces, and electron secondaries from stopped muons and cosmic ray showers with ionizing tracks.  A known intrinsic isotopic contaminate in the graphene target
is the $^{14}$C $\beta$-decay.  The rate of $^{14}$C $\beta$ electrons in the signal window can be
as large as $10^6$ times the neutrino capture signal depending on the source of carbon used to fabricate the
graphene.  Underground source of hydrocarbons and carbon dioxide have been used in the Borexino experiment
with levels of $^{14}$C of $10^{-18}$, giving over an order of magnitude lower levels than required for PTOLEMY.

With a planar surface of 100m$^2$, the cosmic ray flux on the tritium target is roughly 20kHz.  
There is a low probability for cosmic ray interactions to directly liberate low energy electrons into the magnetic
aperture of the PTOLEMY spectrometer.  The probability for direct interactions with the target support material
can be minimized, but ultimately, for an experiment located on or near the surface, the required background
rejection factor is enormous.  For PTOLEMY, a background rejection factor of order 10$^{12}$ is needed.  A narrow
energy window of 0.5eV at the tritium endpoint is estimated to provide a suppression of roughly 10$^7$--10$^8$.
The remaining factor will require either moving the experiment underground or a highly efficient muon veto with hermetic coverage in the region of the tritium target.
Isotopic contaminants generated in cosmic ray showers are a primary source of backgrounds given the long exposure time.
Potential sources of neutrino backgrounds to the relic neutrino capture
process were evaluated for solar neutrinos and found to be negligible below 100keV.  Nuclear reactors and free neutron decay produce primarily electron anti-neutrinos and will therefore not capture on tritium.
The primary source of reducible background is from tritium $\beta$-decay.  Techniques for the further narrowing of the electron energy resolution beyond the calorimeter-only measurement can
provide additional rejection for this background.
The analysis of the tritium endpoint is presented in section~\ref{sec:physics}.

%% file: sections/Performance/performance.tex
\vspace{-0.25 truein}
\section{Physics performance studies\label{sec:physics}}

\begin{figure}[h!]
\begin{center}
\includegraphics[width=0.4\textwidth]{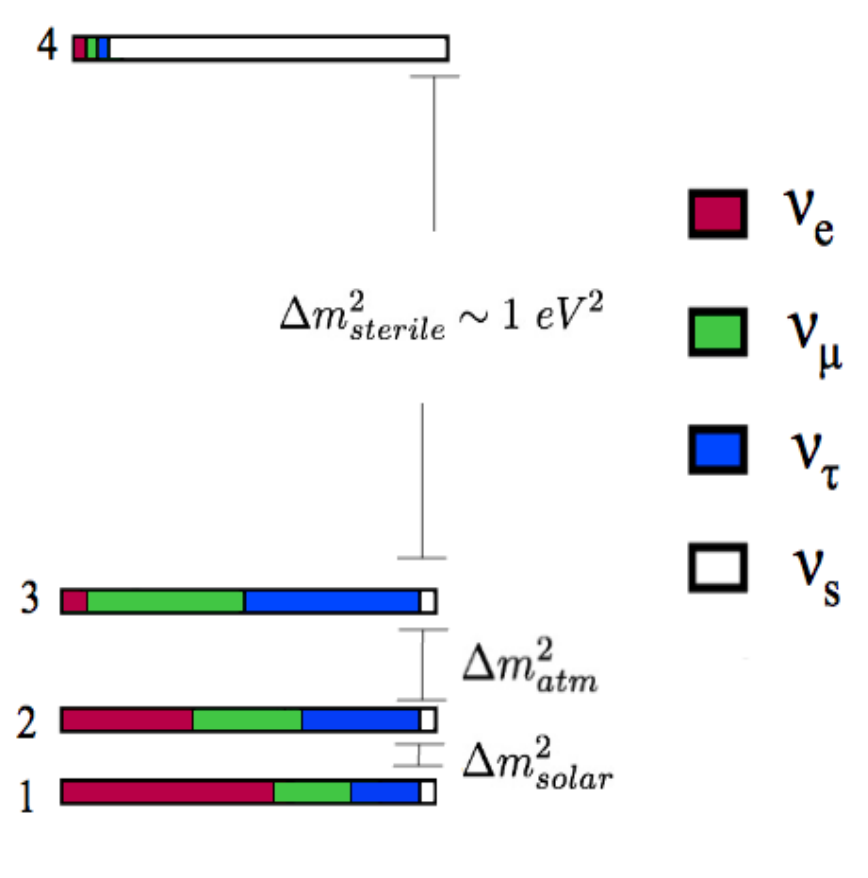}
\caption{An example sterile neutrino addition to the three-family neutrino mass spectrum, compatible with current oscillation data.  This is shown for a sterile neutrino mass of 1~eV.  A small admixture of electron flavor in the sterile
neutrino, represented by the heaviest mass eigenstate, would enable the tritium neutrino capture process and produce an energetic electron signal in the PTOLEMY experiment.  For non-relativistic, relic, sterile neutrinos,
the electron spectrum would be a narrow peak located above the tritium endpoint at the endpoint energy plus
the mass of the heaviest mass eigenstate.
\label{fig:sterilemassspec}}
\end{center}
\end{figure}

The primary design goal of the PTOLEMY project is to achieve direct detection of the 
cosmic relic neutrinos.  Big Bang cosmology predicts an average number density of 56 electron
neutrinos per cubic centimeter everywhere in space and modifications to this prediction from 
gravitational clustering increase the local number by a factor 1-100 depending on the neutrino mass.
To achieve the signal sensitivity needed to detect the relic neutrino density, the measurement
precision of the endpoint electrons must be pushed to the level of the neutrino mass and the
tritium target mass needs to achieve 100gram-year exposures.  In advance of achieving sensitivity
to the relic neutrino density, there are important contributions that can be made to the indirect 
measurement of the neutrino mass.  The sub-eV measurement capability of the calorimeter is
a fundamentally new approach to the neutrino mass measurement.
Similarly, PTOLEMY is capable of completing the 
highest sensitivity search for keV-scale dark matter candidates in the form of sterile neutrinos with an 
admixture of electron flavor, at the level $|U_{e4}|^2$ of $10^{-4}$--$10^{-6}$, depending on the 
sterile neutrino mass.  
An example sterile neutrino addition to the three-family neutrino mass spectrum, compatible with current oscillation data, is shown in Figure~\ref{fig:sterilemassspec}. 
Beyond these benchmark physics processes, PTOLEMY has unique
sensitivity to a weakly interacting component of the matter composition of the Universe with the 
potential to uncover unexpected properties of early universe evolution through anomalous 
contributions to the neutrino capture energy spectrum and in the understanding of elementary particle
physics, such as a finite neutrino lifetime.

The physics performance for relic neutrino detection, neutrino mass measurements, and dark
matter searches are evaluated with corresponding background models and systematics related
to those measurements.  Many of the backgrounds are common to all measurements and will be
discussed separately.  Specific backgrounds related to each measurement are discussed in the analysis
sections.  Detector performance benchmarks are specified in the discussion of analysis sensitivity
for each of the physics goals.











\subsection{Background sources and data-driven estimation techniques}

The GEANT4 simulation package is used to model the target geometry, cryostat, and calorimeter response.
These simulation are used to evaluate the signal efficiency for endpoint electrons to enter the
aperature of the MAC-E filter.  The simulation models the MAC-E filter cut-off precision and the
flux of $\beta$-decay electrons from tritium and $^{14}$C on the calorimeter.
The effects of pileup on the calorimeter readout are emulated and the digital data are fed into 
reconstruction algorithms.
Background sources are generated and tracked with GEANT4.  Background particle sources are those that
scatter or capture on the material components in the experimental volume or decay in flight to produce 
secondary electrons.  The background electron sources are propagated through the MAC-E magnetic
filter, RF tracker, and hit the calorimeter.  The resulting measurements enter the endpoint analysis.

The magnetic and electric field maps are computed with COMSOL and input into the GEANT4 simulation 
for charged particle tracking.  COMSOL also has charged particle tracking and provides a detailed 
benchmark to compare with GEANT4.

\begin{figure}[h]
\begin{centering}
\includegraphics[width=0.65\textwidth]{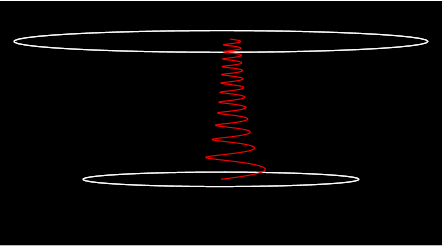}
\par\end{centering}

\caption{Particle tracking of 10keV electron from tritium decay in the GEANT4 simulation of simplified
PTOLEMY detector.  The cyclotron motion in the expanding magnetic flux is clearly visible.
\label{fig:geant4}}
\end{figure}

Cosmic ray background measurements with a muon coincidence have logged data in the PTOLEMY test cell
basement where the prototype is located.  Approximately 18ft. of concrete overburden reduces the cosmic 
ray flux by a roughly a factor of 2 relative to the surface.  A preliminary analysis of the radioistopic 
backgrounds, cosmic ray induced capture processes, and other direct background sources are being measured
with a windlowless silicon APD within the MAC-E magnetic field in a roughly 10$^{-6}$~Torr vacuum.  
A graphite disk with a calibrated $^{14}$C source strength will provide a wide energy spectrum of
$\beta$-decays and the graphite will act as a scattering target for secondary electron production
from background sources.

%% file: sections/Tritium/tritium.tex
\vspace{-0.25 truein}
\section{Tritium target}

PPPL is well positioned to support PTOLEMY tritium requirements.  The PPPL site has already demonstrated the ability to safely handle 100 grams of tritium during the operation of the Tokamak Fusion Test Reactor (TFTR).  PPPL Tritium Facility, located at D-Site, contains the appropriate equipment for supporting prototype, mid-range, and full scale PTOLEMY configurations. 
The facility is fully instrumented for storage, purification, analysis, delivery, and recovery of tritium.  The PPPL Tritium Facility was the first in the United States to successfully introduce tritium into a magnetic fusion energy (MFE) device.  Still to this date, PPPL Tritium Facility holds the world record for successfully fueling more than 1,000 D-T shots in TFTR.

The PPPL tritium system includes a 1,000 CFM tritium recovery system designed to maintain negative pressure on large volumes (consistent with full scale PTOLEMY chamber) in the event of a chamber leak. This tritium recovery device is one of several safety systems available at D-Site for use during prototype, mid-range, and full scale PTOLEMY.
As a result of previous tritium operations, in support of the D-T program at the lab, PPPL has experience handling tritiated graphite and is currently investigating the production of tritiated graphene.  This experience coupled with a newly commissioned plasma based nano-technology laboratory provide for a strong technical platform for undertaking this work. 
Additionally PPPL maintains comprehensive analytical and engineering capabilities which include machine shops for the fabrication of high tolerance components. The laboratory also maintains Industrial hygiene, radiological control, and emergency services staffs.

Surface-deposition tritium sources were studied extensively
by the Mainz experiment [arXiv:hep-ex/0412056v2 6Jan 2005].
In the Mainz neutrino mass experiment, a highly oriented pyrolytic 
graphite (HOPG) substrate cooled to 1.85K held approximately 140 
layers of shock condensed, amorphous T$_2$ on a surface area 
of 2~cm$^2$.
The advantages of this source, as described by the Mainz
experiment, were the (i) low backscattering rate, (ii) atomic
flat surface, (iii) high purity.  The substrate was cooled to below 2K
to stop the process of dewetting of the T$_2$ which subsequently led
to the build up of thick T$_2$ crystals on the surface.
The substrate was glued to the copperhead of the 
cryostat with a silver-loaded, heat conducting glue in order to establish
strong thermal and electrical contact.  The electrical charging of the weakly conducting
T$_2$ atomic layers from the process of tritium $\beta$-decay resulted in a 20mV/layer 
slope on the voltage reference for the tritium $\beta$ energies.  The self-charging versus 
thickness and the time-dependent build up of H$_2$ contaminant in the vacuum system 
on the top surface of the tritium substrate were important systematics on the upper limit
placed by the Mainz experiment on electron-flavor neutrino masses.

Important studies were conducted on the properties of surface-deposition tritium
sources and compared with gaseous sources in the context of the Mainz 
experiment [Otten, DOI 10.1007/s10751-009-0150-2 and 
Aseev et al., Eur. Phys. J. D 10, 39{52 (2000)].  The total inelastic cross-section
was measured to be $3 \times 10^{-18}$cm$^2$ for $E=18.6$keV with a peak
energy loss per scatter of approximately 12--14eV and an average energy loss
per scatter of approximately 30eV.  The total inelastic cross-section was measured
with an electron beam for gaseous T$_2$ and with $^{83m}$Kr for films of D$_2$
prepared in a similar way as the Mainz T$_2$ shock condensed source.
The total inelastic cross section effectively limits the column density of T$_2$
to $5 \times 10^{17}$ molecules/cm$^2$ before single and multiple scattering
in the source begins to dominate the energy resolution of the tritium $\beta$ at
$E=18.6$keV.

The $\beta$ decay of tritium starting from a tightly bound molecular state, T$_2$,
introduces a range of final states and final-state energies of the non-decay T, 
$^3$He, two atomic electrons, and the $\beta$-decay electron.  The binding energy of T$_2$
is approximately 4.6eV.  A number of closely-spaces ro-vibrational states can be excited in the
final state ($^3$He--T)$^+$ ion and induces a smearing of approximately 0.5eV on
the neutrino mass measurement.  The energy range of these low energy states
extends up to 4eV with continuum $^3$He+T production above 1.9eV.  There is a
10eV gap between the low-energy ro-vibrational states and the more energetic final 
states of the T$_2$ $\beta$-decay.  For gaseous T$_2$ sources at a temperature of
27K, the Doppler shift broadening is predicted to be of order 0.2eV [M. Heine, MIT
Thesis 2008].  Doppler broadening shifts in surface-deposition sources are of a similar
magnitude and can be further suppressed as a linear function of operating temperature.

The PTOLEMY experiment is based on an atomic thickness tritium target.  The MAC-E filter geometry 
restricts the trajectories of electrons
emitted from tritium to within the cyclotron radius of the adiabatically varying 
magnetic field lines threading the tritium target surface.
The large surface area required for relic neutrino detector can be factorized into isolated
tritium target areas using a method of magnetic ducts described in the section on the MAC-E filter.
Half of the $4\pi$ solid angle of the isotropically emitted electrons from a planar tritium 
target are accelerated with a precision voltage reference into a high magnetic field region 
at the opening of the MAC-E filter.  
The tritium target surface can be over an order of magnitude larger than the bore area
of the MAC-E filter magnet using this acceleration technique. 

The original Mainz method of highly oriented pyrolytic 
graphite (HOPG) substrates has a natural extension in the area of graphene substrates.
Single atomic layers of graphene hold hydrogen through chemical absorption with a measured
desorption energy temperature dependent slope of 19.6kJ/mol, corresponding to approximately
0.42eV binding energy for hydrogen.  This measurement is being repeated with tritium on graphene.
The chemical absorption process places individual tritium atoms in the $sp$-3 orbital
of the $^{12}$C nuclei in alternating sites on the top and bottom sides of the graphene layer,
as shown in Figure~\ref{fig:GrapheneH} for the example of a graphene nanotube.  
The binding energy of tritium on graphene is an order of magnitude weaker than the T$_2$ binding 
energy.  The differential density of final states of the $^3$He is expected to be reduced by an order
of magnitude.  By binding the tritium to individual layers of graphene, the high conductivity of graphene 
is also expected to eliminate the effect of charging and voltage reference shifting observed by the
Mainz experiment.  The conductivity of graphene is reduced by the hydration of the graphene,
and therefore the uniformity of the voltage reference will need to be studied for fully and partially
hydrated graphene.

The technology of graphene-held atomic tritium is in under development
by the PPPL nanotechnology lab for use in the PTOLEMY experiment.
Hydrogen plasma temperatures below 1eV have been achieved for the hydration of graphene.
The method of cold plasma fabrication of graphene sheets creates a substrate of a single atomic layer.
Existing sample of tritium held on graphite tiles are being examined to identify whether the
predicted chemical absorption binding mechanism is observed for tritium.  Similarly, the absence
of $^3$He bound to the graphene surface would provide important input on the possible final states of 
graphene-held tritium $\beta$ decay.
The spectrum of excited states generated by an endpoint electron emission from a tritium atom bound 
to graphene to the $^3$He final state is under numerical study.  Given the narrow phase space 
of the decay process near the endpoint, the population of the excited states has an important impact on 
the shape of the $\beta$-decay spectrum.  A similar evaluation of the neutrino capture signal
process is under study.

\begin{figure}[h!]
\begin{center}
\includegraphics[width=0.3\textwidth]{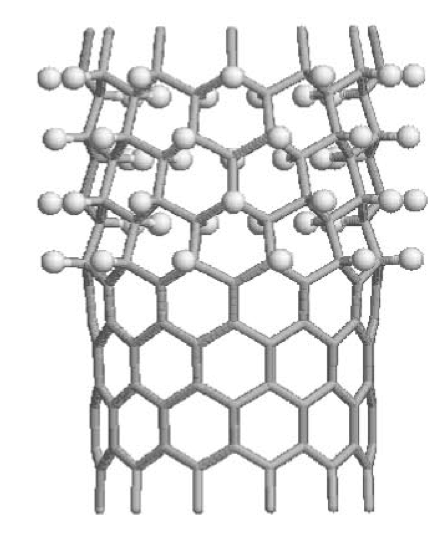}
\caption{The hydration of graphene places an individual hydrogen isotope at every carbon
site, alternating front and back layers.
\label{fig:GrapheneH}}
\end{center}
\end{figure}

The challenges of developing the PTOLEMY tritium target are common to many aspects of hydrogen
fuel cell development.  The graphene-bound tritium target would hold 20\% by weight of tritium
and be stable at room temperature.  The reusability of the graphene for tritium storage and the spacing
of graphene layer to allow efficient flow of tritium in and out of the graphene structure is common
to hydrogen fuel cell design.  The PTOLEMY target design will incorporate the magnetic
geometry to efficiently capture electrons from the tritium source into the aperture of the MAC-E filter.
The change in conductivity from the insulator to conductor transition of graphane to graphene can be been
used to monitor hydration levels of graphene and in the context of PTOLEMY could be used to 
monitor the tritium decays from individual graphene substrates.  Fine granularity instrumentation of the tritium
target with sensitivity to the $^3$He recoil energy and pulse timing has the potential of reducing the hit rate
down to the level where time-of-flight information for the entire endpoint electron trajectory could be
measured to a precision of 10$^{-4}$ for trajectories lasting more than 10$\mu$s from target to calorimeter.
Initial evaluation tests of the graphene-T insulator to graphene conductor transition will be performed
with a flux-locked loop.  The graphene conductivity is expected to increase in steps with each tritium
decay as the liberated hydrogen isotope frees up a conduction electron to the graphene substrate.
The trend of the conductivity increase is expected to be a measure of the number of tritium decays.

An experimental program of evaluating the excitation spectrum, binding
energies, and electron scattering effects in the tritium target are planned 
using the PTOLEMY prototype.
An electron-gun calibration system will provide
in situ energy calibration for the MAC-E filter, RF tracker, cryogenic 
calorimeter, and time-of-flight system.  Similarly, in situ surface scattering
data on a target will be analyzed using the full spectrometer.
The continued investigation of high resolution tritium targets is of general interest
to the entire neutrino mass community.
PTOLEMY aims to develop an atomic tritium target weakly bound to graphene with
significantly lower endpoint electron energy smearing than diatomic T$_2$.

\subsection{Graphene Fabrication, Tritration, and Surface Analysis}

Graphene is fabricated at PPPL with cold magnetized plasmas at low pressure.  A graphene
substrate is then exposed to a cold magnetized plasma of tritium with $T_e<1$~eV to produce
the graphene-held tritium target material for PTOLEMY.  Samples of tritium-exposed graphene
samples are under analysis at the Princeton Institute for the Science and Techology of Materials (PRISM)
and at the Savannah River National Laboratory (SNRL).

\begin{figure}[h!]
\begin{center}
\includegraphics[width=0.65\textwidth]{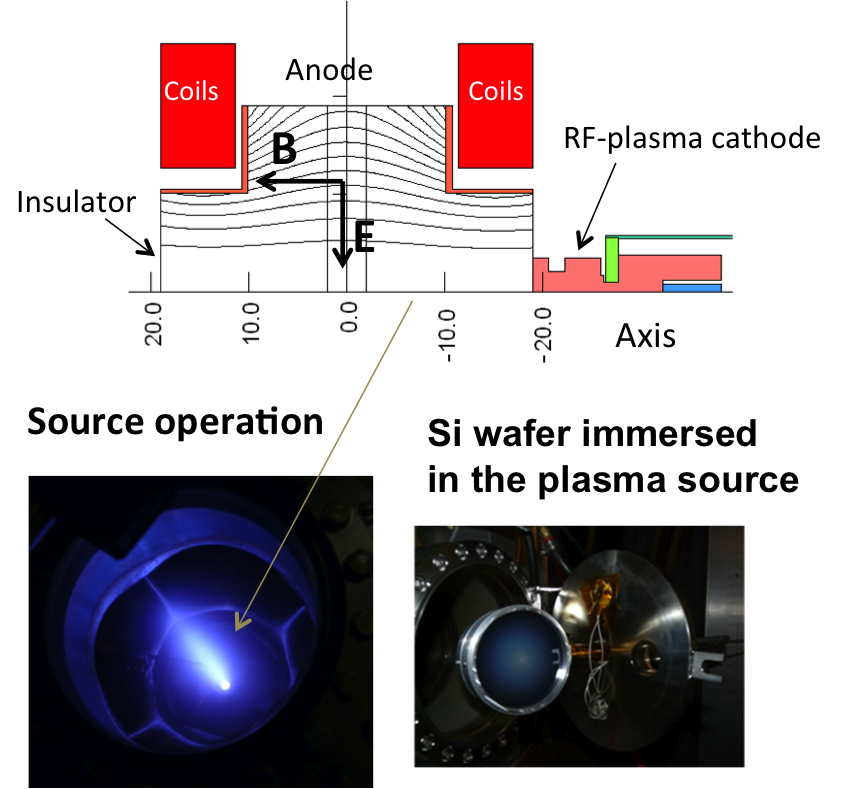}
\caption{Plasma-driven graphene fabrication infrastructure at PPPL.
\label{fig:GrapheneFab}}
\end{center}
\end{figure}

\subsection{Large Area Tritium Target Design}

The sensitivity of PTOLEMY to measure the relic neutrino density is set by the total tritium target area and total
tritium mass, the exposure time, and the product of the acceptance times efficiency for endpoint electrons to enter the MAC-E filter without scattering or other potential multi-eV energy smearing effects.  Sources for energy smearing relating to the binding energy of atomic tritium to the graphene and of the final state energy spread of the $^3$He are addressed in the graphene-T material analyses.  The intrinsic limitations of electron scattering, the target area, and acceptance times efficiency for endpoint electrons come from the design of the tritium target interface to the MAC-E filter.  The approach taken here is to use a monolayer of graphene with one atomic tritium bound to each carbon
atomic with a weak sub-eV binding energy as is characteristic of the graphene structure.  The exposed
area of the graphene is taken to be normal to the magnetic field lines in the tritium storage containers.
The magnetic field lines from the tritium storage containers are pulled into magnetic ducts that feed the
MAC-E filter and subsequently the RF tracker and calorimeter systems.  These flux lines are closed by returning
them to the entry point of the tritium storage container through a flux return loop.  The challenge for the tritium
target source is to expose a large area of graphene to the magnetic field lines.

In the simplest geometry, the graphene substrate spans the cross sectional area of the tritium storage container.
This defines the total area per storage container, and this area times the number of storage containers feeding
the MAC-E filter define the total tritium target area.  The acceptance for the endpoint electrons starts from the
$2\pi$ solid angle of the graphene surface normally oriented to the magnetic field.  Then, losses on the acceptance
times efficiency can occur from several sources.  If the MAC-E filter cut-off is not low enough or sharp enough,
then large opening angles of electrons originating from the graphene surface will not make it through the
filter, rejected through a magnetic mirror bounce before entering the RF tracker.  If the transverse kinetic energy
is not large enough, then the RF tracker will not be able to detect the single electron signature - and though these
electrons can still be recorded by their subsequent interaction with the calorimeter, the non-electron backgrounds
to this class of electrons will be substantial.  The endpoint electrons can also lose in efficiency if they scatter
with gas in the vacuum or if they are inefficiently detected by the calorimeter - either due to geometric efficiency,
or dead-time or other instrumental effects that degrade the energy resolution of the endpoint electron to the point
where the $\beta$-decay backgrounds from tritium dominate.

In a large area geometry, a series of monoatomic graphene area elements use electrodes to drift endpoint electrons out of the magnetic lines that directly thread the graphene surface and into a shared flux that enters the magnetic ducts leading to the MAC-E filter.  The drifting of the gryo center of the endpoint electron cyclotron motion is 
achieved with an ${\bf E} \times {\bf B}$ field configuration leading to a non-relativistic drift velocity of 
${\bf v}_{drift} = ({\bf E} \times {\bf B})/B^2$.  The drifting of the gyro center stops when the electron drifts out of the
electric field region and enters a set of shared magnetic flux lines that feed the MAC-E filter.  A repeated geometry
for scaling up the total area of the graphene is shown in Figure~\ref{fig:tritiumlargeareageo}.  The simulation
of this structure is done with COMSOL.

\begin{figure}[h!]
\begin{center}
\includegraphics[width=0.65\textwidth]{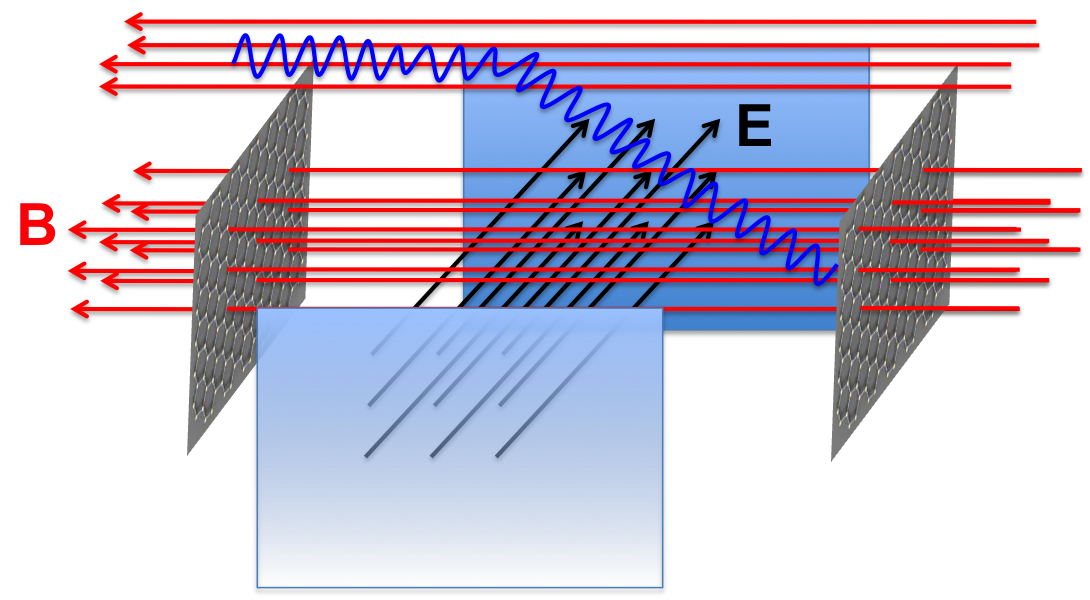}
\caption{The large area tritium target uses an ${\bf E} \times {\bf B}$ field configuration to drift endpoint electrons
into a set of shared magnetic flux lines that feed the MAC-E filter.
\label{fig:tritiumlargeareageo}}
\end{center}
\end{figure}

%% file: sections/MAC-E/mac-e.tex
\vspace{-0.25 truein}
\section{MAC-E filter}

We need a spectrometer with high luminosity and low background to
provide a useful measurement of the spectrum near the endpoint: these
requirements are fulfilled by a MAC-E filter. This type of spectrometer
was first proposed by Beamson et al. and has been used to reduce the
upper bound on the neutrino mass by Troitsk, Mainz and KATRIN \cite{Beamson1980Collimating,Lobashev1985Method,Picard1992Solenoid,KATRINcollaboration2001KATRIN}. 

Two superconducting solenoids produce a magnetic field which adiabatically
guides electrons to a detector. In the center of the spectrometer
-- the analyzing plane -- the magnetic field expands and decreases
by several orders of magnitude to its minimum value, $B_{\mbox{min}}$.
As an electron moves from a region of high to low magnetic field,
its cyclotron momentum perpendicular to the field lines is transformed
into longitudinal momentum along the field lines. At the analyzing
plane, we impose an electrostatic retarding potential, $V$. Only
electrons with a minimum energy parallel to the field lines are able
to pass through, $E^{\parallel}>qV$. Other electrons are reflected
from the barrier. Thus, the MAC-E filter acts as an integrating high-pass
filter. 



At the energies considered in tritium beta decays we may work in the
non-relativistic approximation ($\gamma\leq1.04$). The adiabatic
condition is equivalent to requiring the flux enclosed by the cyclotron
orbit to be constant. In practice, this is fulfilled if the magnetic
field along the circular orbit is slowly changing. As the magnetic
field varies along the trajectory, the cyclotron radius changes to
preserve the constant-flux condition. The adiabatic invariant associated
with this motion is:

\begin{equation}
\phi=\int\vec{B}\cdot\mbox{d}\vec{S}=\mbox{const.}\label{eq:flux}
\end{equation}
We may express the above conserved quantity as: 
\begin{equation}
\mu=\frac{E^{\perp}}{B}=\mbox{const.}\label{eq:mag_moment}
\end{equation}
where $E^{\perp}$ is the electron energy perpendicular to the field
line.

We may use the properties of an ideal MAC-E filter to derive a transmission
function: the total electron energy is conserved, energy contribution
from the magnetic field is negligible, and only the parallel velocity
can be used to overcome the potential barrier. The initial, non-relativistic
energy of an endpoint electron is:
\begin{equation}
E_{0}=\frac{1}{2}mv_{0}^{2}-eV_{0}=E_{0}^{\parallel}+E_{0}^{\perp}-eV_{0}=\frac{1}{2}mv_{0}^{2}\left(\cos^{2}\theta_{0}+\sin^{2}\theta_{0}\right)-eV_{0}\label{eq:energy}
\end{equation}
where $m$ is the electron mass, $e$ is the electron charge, $V_{0}$
is the potential at the source and $\theta_{0}$ is the initial angle
between the electron's momentum and the guiding magnetic field line.
We then use conservation of energy and the adiabatic condition to
express this at any point in the spectrometer:
\begin{equation}
E_{0}=E^{\parallel}+E^{\perp}-eV=E^{\parallel}+\frac{B}{B_{0}}E_{0}^{\perp}-eV.\label{eq:energy_2}
\end{equation}

We are really interested in whether the electron can pass the midpoint
and enter the detector solenoid. For this, we need the value of $E^{\parallel}$:

\begin{eqnarray}
E^{\parallel} & = & \frac{1}{2}mv_{0}^{2}-\frac{B}{B_{0}}E_{0}^{\perp}+e\left(V-V_{0}\right)\nonumber \\
 & = & E_{0}\left(1-\frac{B}{B_{0}}\sin^{2}\theta_{0}\right)+e\left(V-V_{0}\right).\label{eq:E_parallel}
\end{eqnarray}
If $E^{\parallel}=0$, the electron will reverse direction along the
field line and fail to reach the detector. This condition depends
on both the ratio of magnetic fields and the applied potential barrier.
There are two obstacles an electron must overcome: the magnetic mirror
effect and the potential barrier in the analyzing plane. 

The magnetic mirror effect occurs when a charged particle with momentum
at an angle to a magnetic field line moves to a region of higher magnetic
field. Using Equation~\ref{eq:E_parallel} with $E^{\parallel}=0$,
we rearrange for $\theta_{0}$:

\begin{equation}
\theta_{0}=\sin^{-1}\left(\frac{B_{0}}{B}\frac{E_{0}+V-V_{0}}{E_{0}}\right).\label{eq:theta}
\end{equation}
First consider the case where $V=V_{0}$. If the particle moves to
an area of lower magnetic field, there is no limit on its pitch angle
to the field lines. However, if it moves to an area of higher field,
$B>B_{0}$, there exists a maximum initial angle larger than which
an electron will reverse its motion along the field line. To maximize
the count rate we need to maximize the range of electron emission
angles which will reach the detector. This can be achieved by ensuring
the field within the detector solenoid is smaller than or equal to
that within the source solenoid. 

\begin{figure}[h]
\begin{centering}
\includegraphics[width=0.6\textwidth]{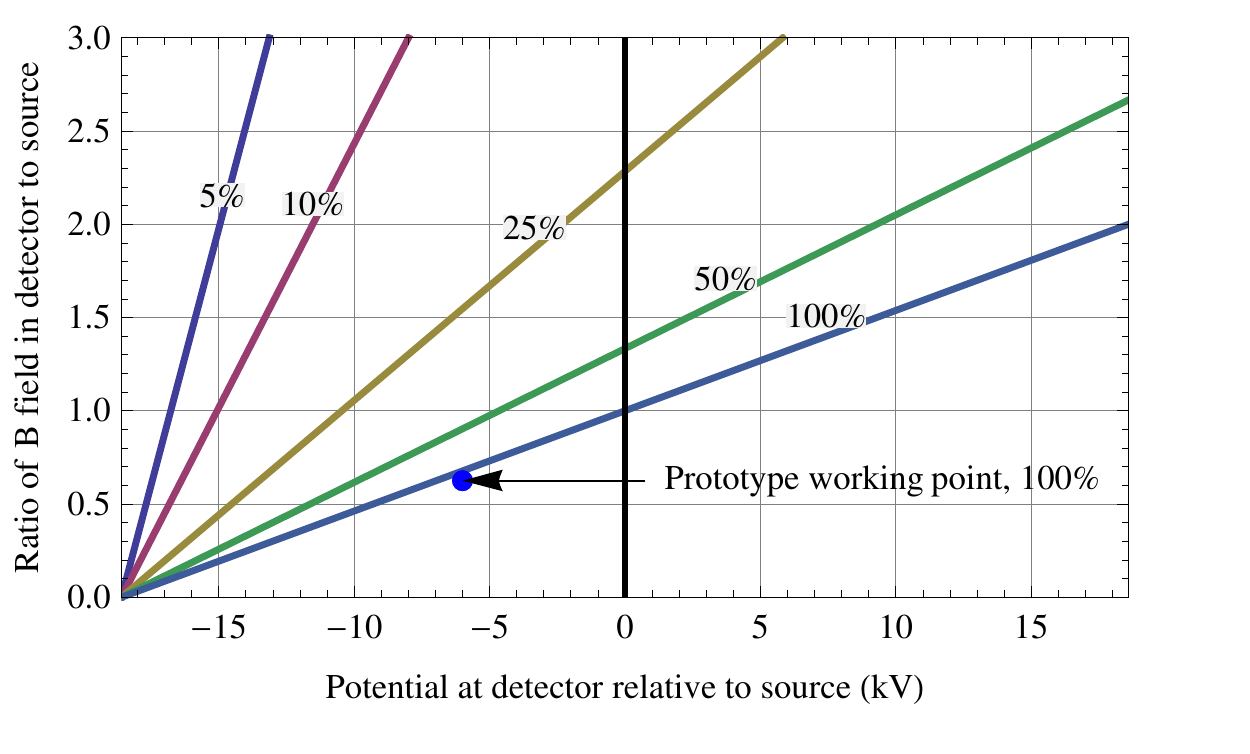}
\par\end{centering}

\caption{Change in capture fraction as potential difference and magnetic field
ratio between source and detector are varied. The lines are labeled
with the percent of electrons that will reach the detector. The working
point for the prototype has been marked.\label{fig:b_vs_v}}
\end{figure}

When the electric potential is also considered, it is possible to
accelerate electrons to compensate for this mirror effect. The consequence
of this for the fraction of emitted electrons able to reach the detector
is shown in Figure~\ref{fig:b_vs_v}.

Next we consider both the magnetic mirror effect and the electrostatic
cutoff together. This gives a transmission function which describes
what fraction of the $2\pi$ solid angle will reach the detector for
a given initial electron energy. The electrostatic cutoff affects
the position and slope of the decrease from 1 to 0 while the magnetic
mirror effect reduces the maximum available solid angle. The transmission
functions are:

\begin{eqnarray}
T\left(E_{0}\right)_{\mbox{mirror}} & = & \frac{\Omega\left(E_{0}\right)}{2\pi}=1-\sqrt{1-\frac{B_{0}}{B_{\mbox{d}}}\frac{E_{0}-V_{0}+V_{\mbox{d}}}{E_{0}}}\label{eq:mirror}\\
T\left(E_{0}\right)_{\mbox{filter}} & = & \frac{\Omega\left(E_{0}\right)}{2\pi}=1-\sqrt{1-\frac{B_{0}}{B_{\mbox{m}}}\frac{E_{0}-V_{0}+V_{\mbox{m}}}{E_{0}}}\label{eq:filter}
\end{eqnarray}
where subscripts indicate initial values ($X_{0}$), values at the
midpoint ($X_{m}$) or values at the detector ($X_{d}$). 

\begin{figure}[h]
\begin{centering}
\includegraphics[width=0.6\textwidth]{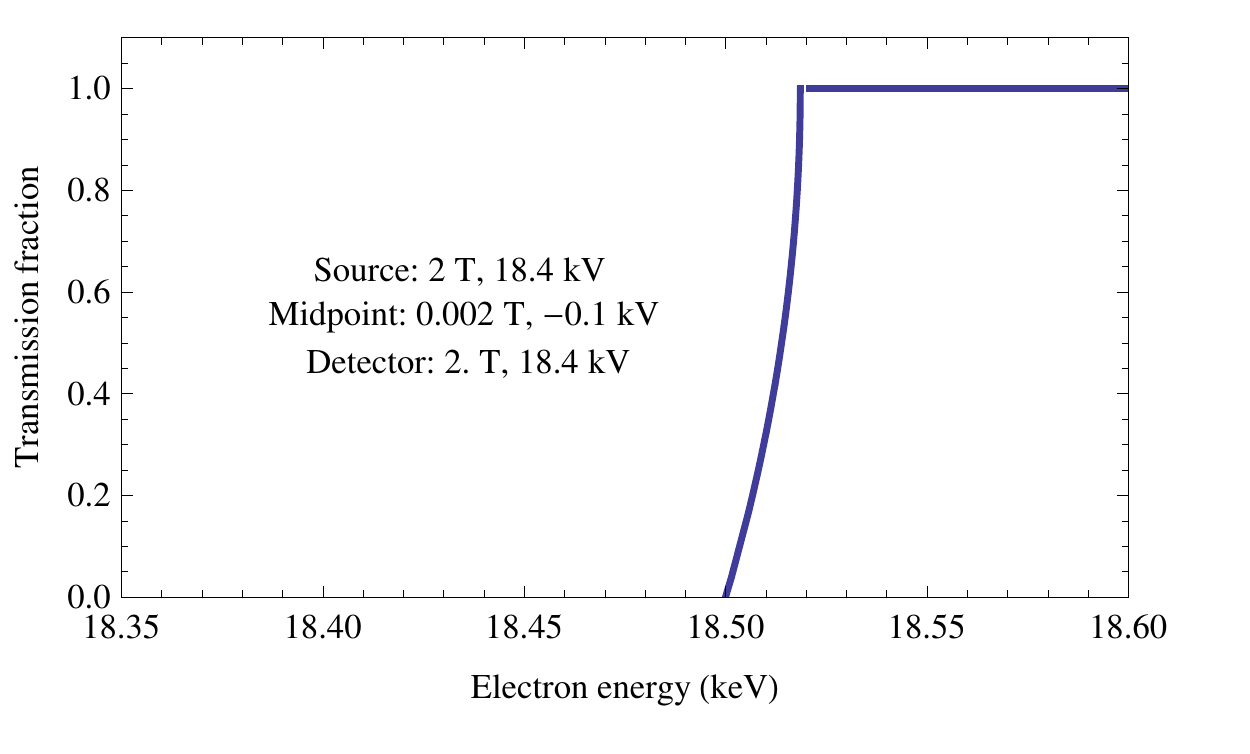}
\par\end{centering}

\caption{An example transmission curve as a function of initial electron energy.\label{fig:transmission}}
\end{figure}

The overall transmission fraction is the minimum value of the two
individual functions. If $T_{\mbox{mirror}}$ allows through 50\%
of the solid angle at a given electron energy while $T_{\mbox{filter}}$
allows only 30\%, the transmission coefficient will be 0.3 -- the
magnetic mirror effect would block particles which do not pass the
filter anyway while leaving those with smaller pitch angles untouched. 
We plot an example transmission function in Figure \ref{fig:transmission}.

For PTOLEMY, the use of a detector with precision calorimetry removes the need
for a sharp cutoff -- the main role of the MAC-E filter is to remove
enough of the spectrum below the endpoint so that the detector may
function without being swamped by large signal rates.

\subsection{Segmented Magnetic Ducts}

\begin{figure}[h]
\begin{centering}
\includegraphics[width=0.25\textwidth]{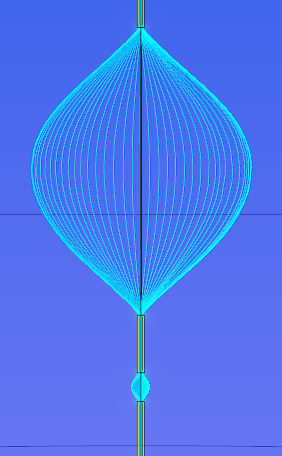}
\includegraphics[width=0.7\textwidth,height=6.7cm]{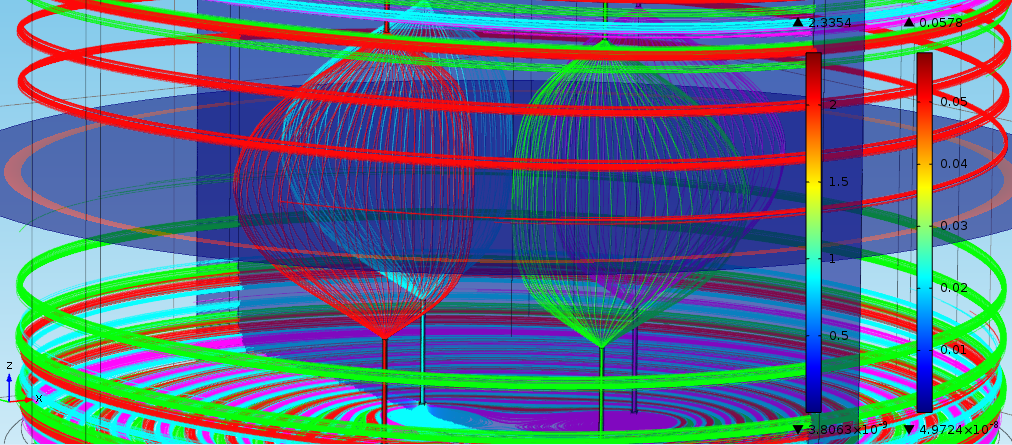}
\par\end{centering}

\caption{The geometry of a single magnetic duct is shown on the left with a small gap to encapsulate an
individual tritium target container on the bottom and a large gap on the top that expands the
magnetic flux lines into a MAC-E filter.  On the right is a view of the field line expansion for four
adjacently placed magnetic ducts feeding a common MAC-E filter volume.  The field lines from each tube 
expand into the space between the gap and close on the originating magnetic duct.
\label{fig:gap_model}}
\end{figure}

\begin{figure}[h]
\begin{centering}
\includegraphics[width=0.65\textwidth]{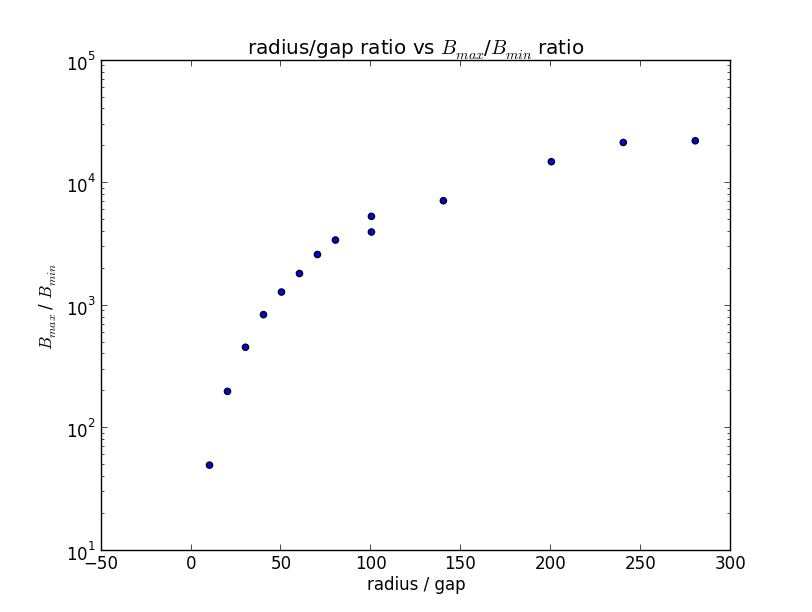}
\par\end{centering}

\caption{The ratio of $B_{max}/B_{min}$ at the center of the gap of a magnetic duct with a radius of 10~cm up to a radius/gap ratio of 100 and with a radius of 5~cm for a radius/gap ratio of 100 and larger.
\label{fig:radius_gap}}
\end{figure}

The electrons from the tritium target follow a unique set of magnetic field lines through the
MAC-E filter, RF tracker, and calorimeter.  This property of the PTOLEMY experiment facilitates
the division of tritium target into separate containers with a system of magnetic ducts, shown
in~\ref{fig:gap_model}.  The size of the gap and the ratio of the gap length to the radius of the duct
changes the value of the ratio, $B_{max}/B_{min}$, of the minimum value of the magnetic field at 
the center of the gap to the magnetic field in the duct.  PTOLEMY is targeting the region of $10^2$--$10^3$
in $B_{max}/B_{min}$ to achieve adequate rate reduction for tritium $\beta$-decay electrons reaching the
calorimeter.  These values are plotted in figure~\ref{fig:radius_gap}
for duct radii of 5cm and 10am.  Off-axis, the minimum B field can drop by an
additional order of magnitude depending on the relative location of neighboring flux tubes.

\subsection{Dome Geometry}

\begin{figure}[h]
\begin{centering}
\includegraphics[width=0.43\textwidth]{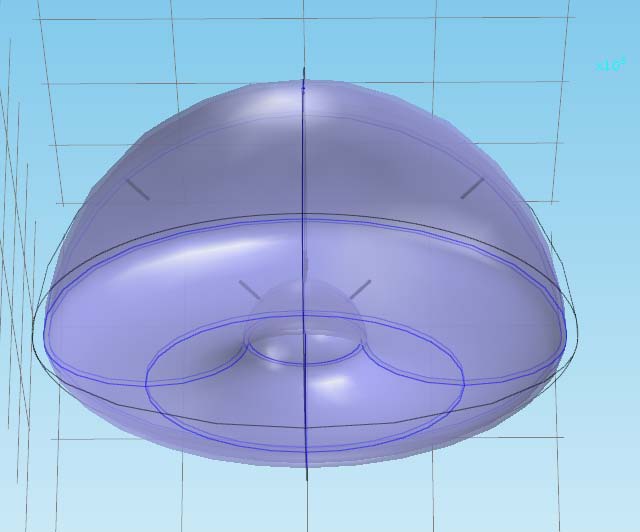}
\includegraphics[width=0.52\textwidth]{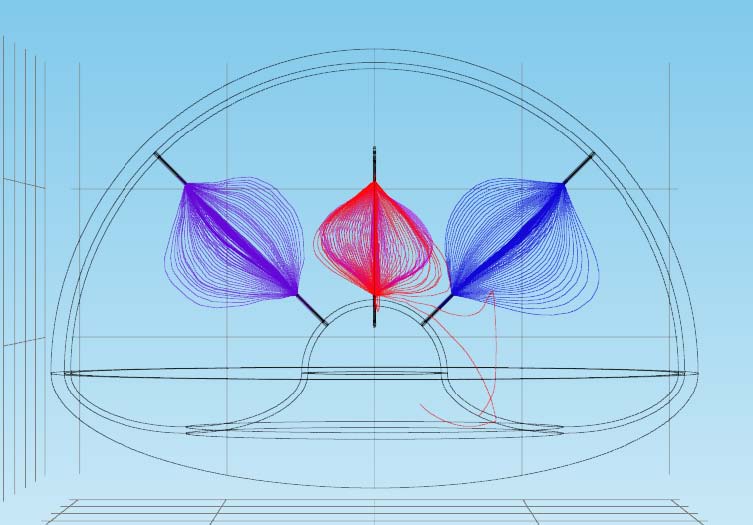}
\caption{A dome geometry for expanding multiple magnetic ducts into a common MAC-E filter vacuum chamber.
On the left is the outer shell of the dome.  On the right are the field lines for a four-quadrant magnetic
duct configuration.
\label{fig:domegeometry}}
\end{centering}
\end{figure}

\begin{figure}[h]
\begin{centering}
\includegraphics[width=0.47\textwidth]{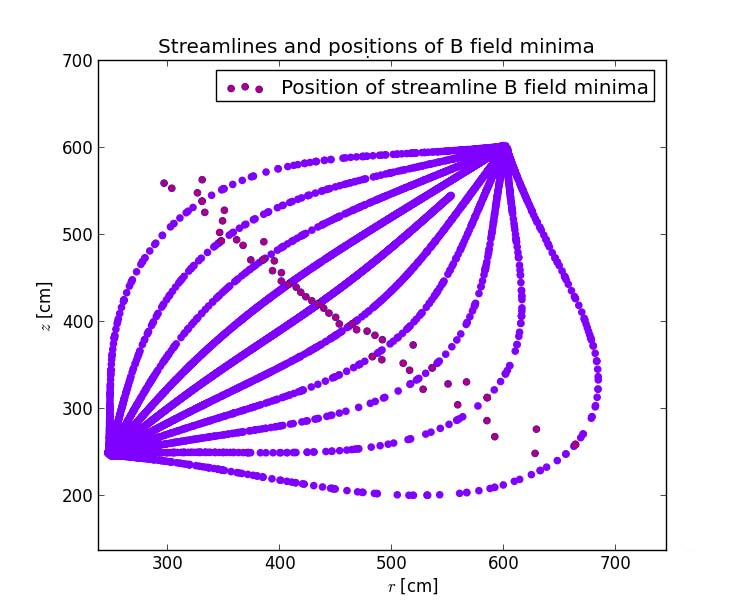}
\includegraphics[width=0.5\textwidth]{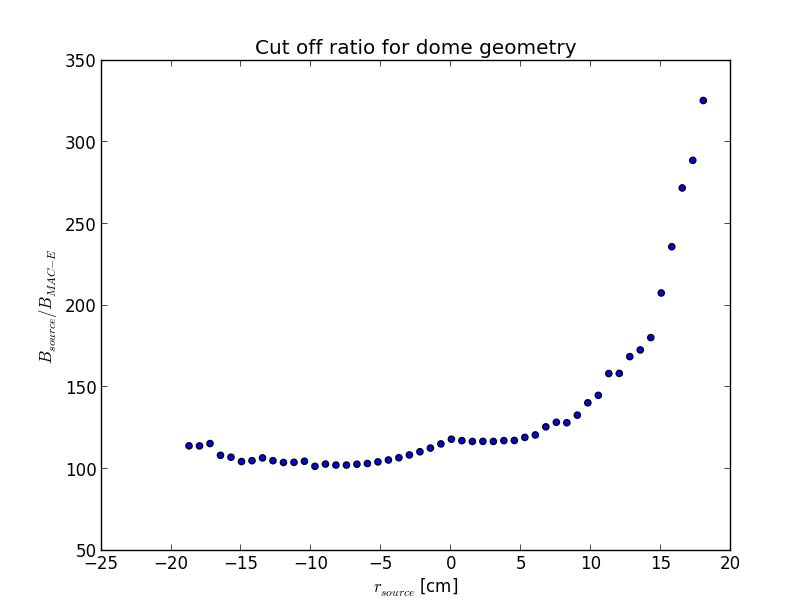}
\caption{The center plane of the dome geometry on the left.  Negative $r_{source}$ is in the direction of 
the central axis of the dome geometry while positive is away.  On the right is the ratio of $B_{max}/B_{min}$ 
over the radius of the tritium target in a four-quadrant configuration, showing a relatively uniform value of roughly 100 for most of the tritium target.  The ratio increases for streamlines that run farthest away from the central axis of the
dome geometry.
\label{fig:domegeoratio}}
\end{centering}
\end{figure}

The non-uniformity of the $B_{max}/B_{min}$ ratio for the MAC-E filter can be minimized by choosing
a packing for the magnetic duct geometry.  The dome geometry, shown in Figure~\ref{fig:domegeometry},
achieves a relatively uniform cut-off for most of the range, which runs from -20~cm for target area radii
that feed streamlines that run near the central axis of the dome and to +20cm for streamlines that
run farthest from the dome axis.  The largest non-uniformity is farthest from
the other magnetic duct field lines.  A large number of closely packed ducts can be made more uniform.
The $B_{max}/B_{min}$ ratio for the dome geometry is shown in Figure~\ref{fig:domegeoratio}.

%% file: sections/RF/rf.tex
\vspace{-0.25 truein}
\section{RF tracker}

\begin{figure}[h!]
\begin{center}
\includegraphics[width=0.45\textwidth]{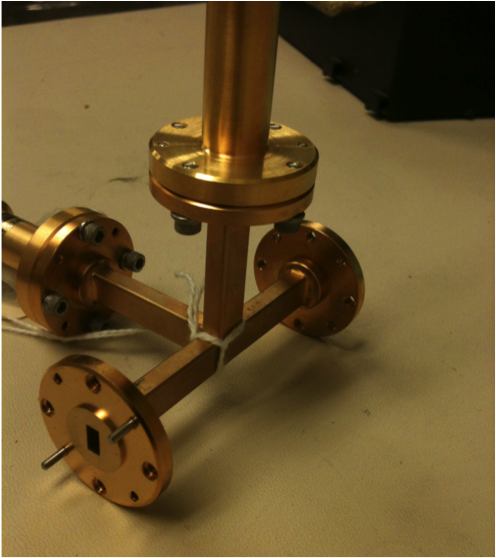}
\includegraphics[width=0.45\textwidth]{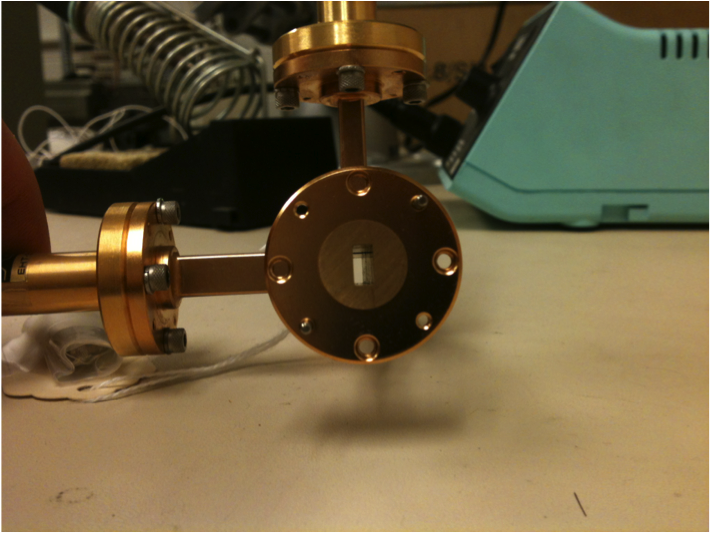}
\caption{A Q-Band (38--46~GHz) ``magic tee" waveguide junction, shown on the left, has two
readout ports that are orthogonal to the straight section to be used as the RF tracker.
An end-on view of the RF tracker waveguide is shown on the right.
\label{fig:MagicTee}}
\end{center}
\end{figure}

A single electron moving in a uniform magnetic field will undergo cyclotron 
motion at a fixed frequency given by
\begin{equation}
f_c = \frac{qB}{2 \pi \gamma m_e c^2}
\end{equation}
where $\gamma$ is the relativistic correction to the total electron mass.
The total power radiated by an electron undergoing cyclotron motion depends
on the cyclotron frequency $f_c$ and the transverse velocity $\beta_\perp$ 
as follows:
\begin{equation}
P_{\rm tot} = \frac{1}{4\pi\epsilon_0} \frac{8 \pi^2 q^2 f_c^2}{3c}\frac{\beta_\perp^2}{1-\beta^2} \ .
\end{equation}
An endpoint electron has an energy of 18.6keV, corresponding to a total
velocity of approximately $\beta=0.26c$.
For $B=1.9$T, the value of $f_c$ is approximately $46$GHz.
An endpoint electron moving transversely to the magnetic field with
$\beta_\perp \approx \beta$ will radiate approximately 
$P_{tot} = 3 \times 10^{-14}$W of coherent RF power.
The coherent up and down cyclotron motion of an electron passing down the
central axis of a matching waveguide will induce a propagating electric 
field in the waveguide.  Figure~\ref{fig:MagicTee} shows the geometry of a Q-band
waveguide in a magic tee configuration.  In this configuration the electron
passes down the axis of the straight section of the magic tee and the two
counter-propagating E-field components induced in the waveguide couple
to the orthogonal legs of the waveguide where a low noise amplifier can
detect the ``RF track'' of the electron.

\begin{figure}[h!]
\begin{center}
\includegraphics[width=0.65\textwidth]{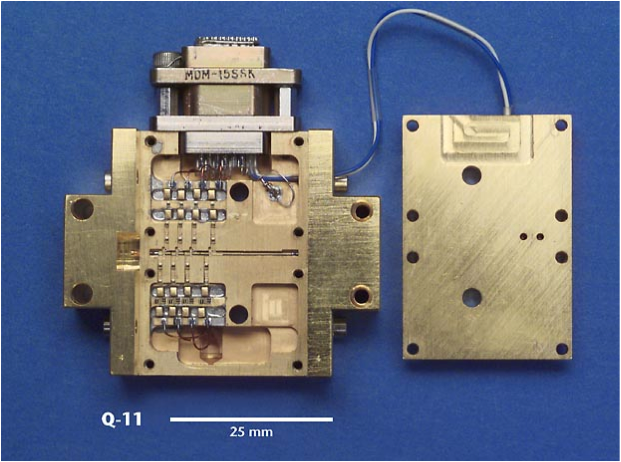}
\caption{A Q-Band (38--46~GHz) WMAP amplifier with a quasilinear noise performance
of approximately 38--46K when operated at a temperature of 20K.
\label{fig:RFAmp}}
\end{center}
\end{figure}

The narrowness of the frequency band around the central cyclotron frequency
sets the total noise of the amplifier.  The width of the band is limited
by the inverse of the transit time of the electron, where for transit times
below 1$\mu$sec the full RF signal is spread over a minimum bandwidth of 1MHz.
The corresponding amplifier noise temperature is set by
\begin{equation}
P_{amp} = k_{\rm B} T_{amp} \Delta f
\end{equation}
and gives approximately $P_{amp} = 8 \times 10^{-16}$W for an amplifier
with a 45K noise temperature and a bandwidth of 1MHz.  The WMAP Q-band 
amplifier, shown in Figure~\ref{fig:RFAmp}, has
a quasi-linear 1K/GHz frequency-dependence on the noise temperature within
the Q-band range (38-46GHz).
The magnitude of $\beta_\perp$, and therefore the total radiation rate, is can be increased 
with the accelerating potential going into a large $B$ field region.  A factor of 4 increase
of radiative signal can be achieved with only 80--90~keV of acceleration.

\begin{figure}[h!]
\begin{center}
\includegraphics[width=0.65\textwidth]{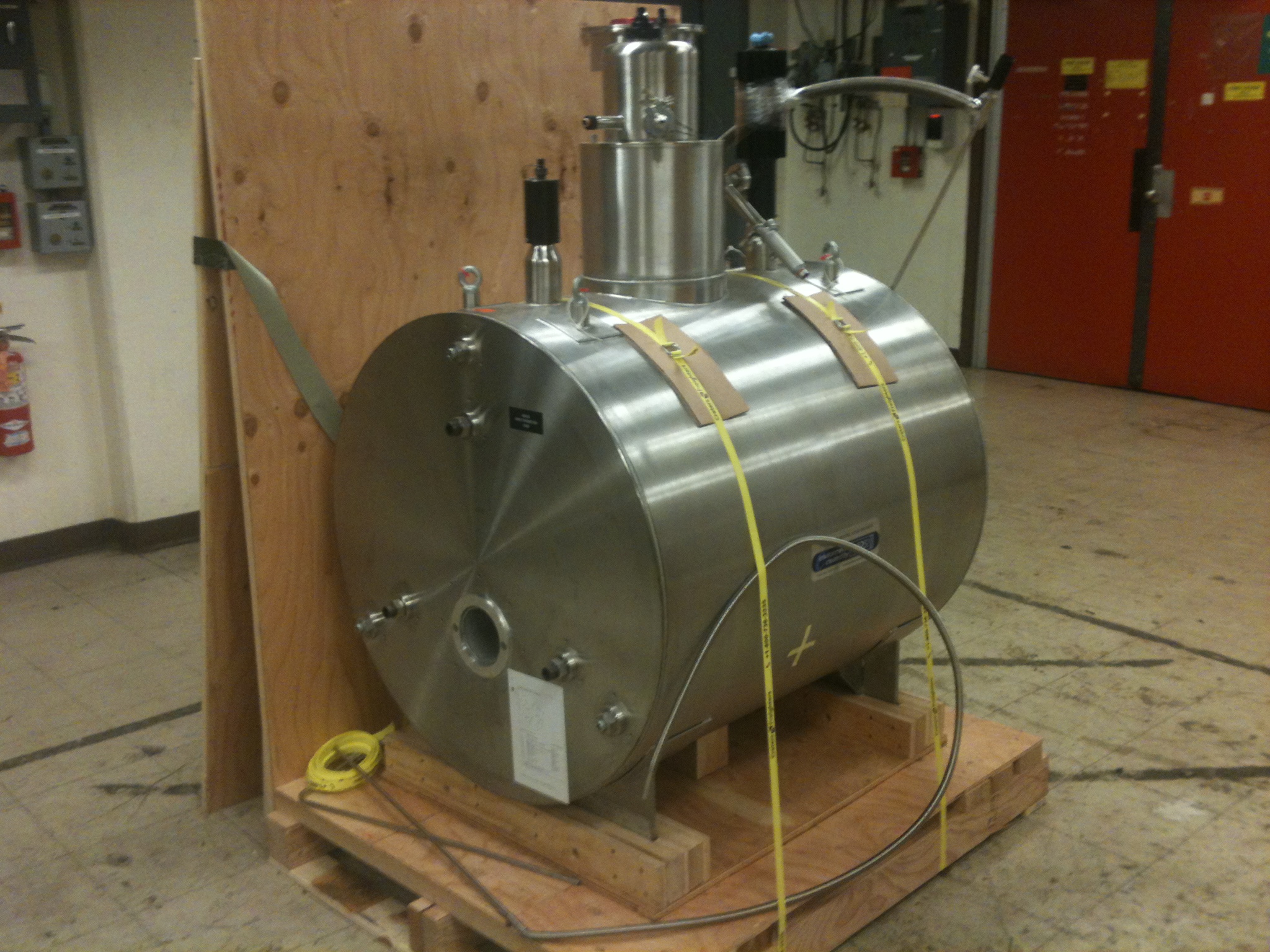}
\caption{7T Ion Cyclotron Resonance (ICR) superconducting magnet for the RF detection
of electron cyclotron radiation in PTOLEMY.
\label{fig:RFmagnet}}
\end{center}
\end{figure}

The Q-band waveguide will be inserted into a 1-meter long uniform magnetic field from
an Ion Cyclotron Resonance (ICR) superconducting magnet, shown in figure~\ref{fig:RFmagnet}.

%% file: sections/Calorimeter/calorimeter.tex
\vspace{-0.25 truein}
\section{Cryogenic Calorimeter}

High precision cryogenic microcalorimetry is central to the design of PTOLEMY.  The estimated electron energy resolution at 100~eV is 0.15~eV.  The energy resolution of the calorimeter is comparable to the upper and lower
limits of the heaviest neutrino mass eigenstate and provides the unique opportunity to directly measure
the energy spectrum of the tritium endpoint over a range of approximately 100~eV.  The final energy resolution
of the endpoint electron is a combination of the calibrated energy measurement from the calorimeter and included
in the calibration the voltage offsets that decelerate the endpoint electron down to approximately 100~eV, energy loss and time of flight corrections that come from the RF tracking measurement, and potentially additional time of flight information from target conductivity monitoring.  The possibility to introduce a momentum spectrometer in front
of the calorimeter for a high granularity deflection measurement is also being considered.

\begin{figure}[h!]
\begin{center}
\includegraphics[width=0.85\textwidth]{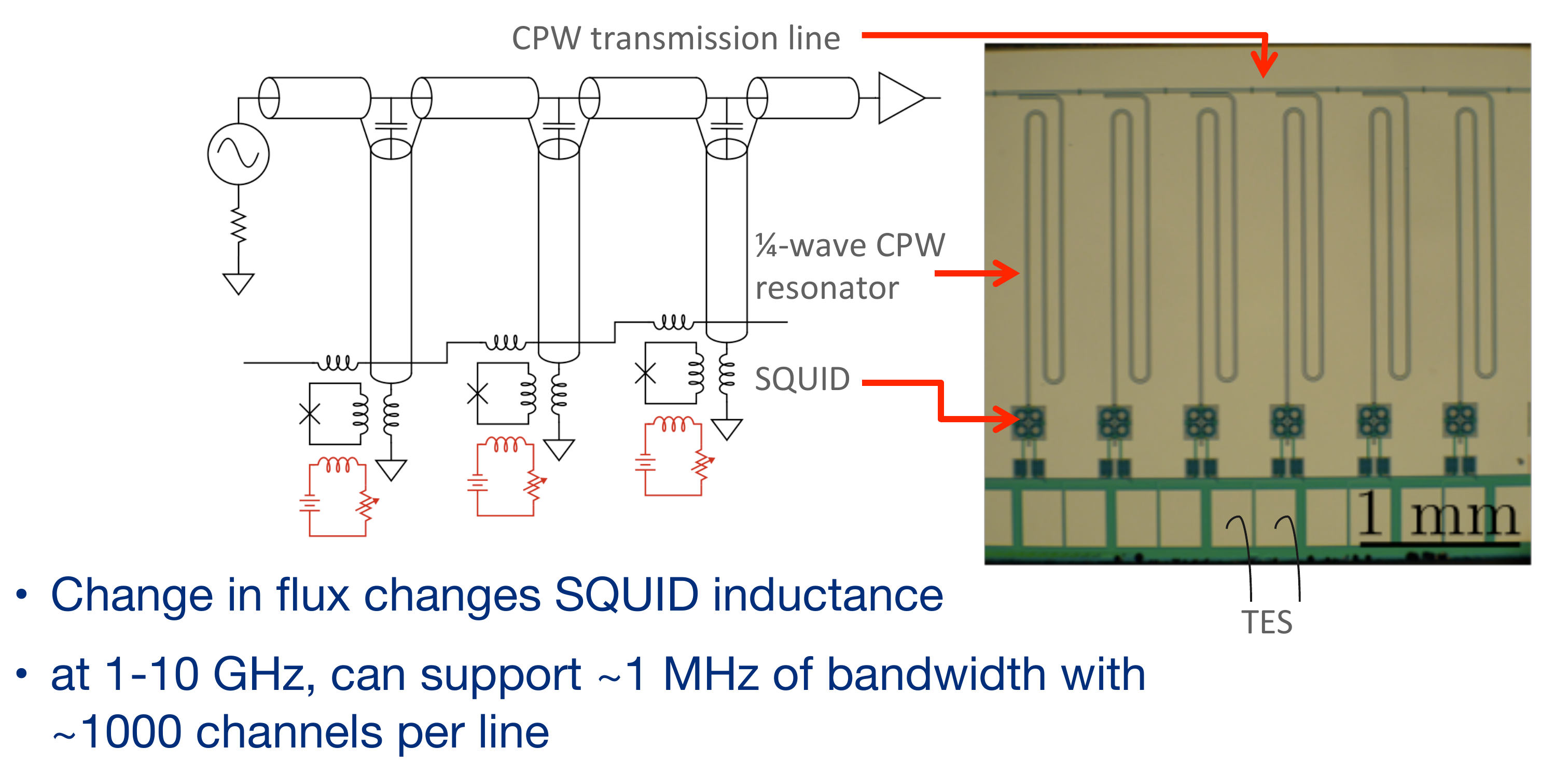}
\caption{Microwave-readout Massive SQUID Multiplexer concept for approximately 1000 channels
per coplanar waveguide(CPW) transmission line operating at 1-10~GHz with a bandwidth of 1~MHz
per channel.  Multiplexers based on this principle have been developed by the NIST/University of Colorado 
TES X-ray microcalorimeter readout.
\label{fig:SQUIDMultiplexer}}
\end{center}
\end{figure}

The response time of an individual calorimeter cell is inversely related to the thermal conductivity, $G$, of the transition-edge sensor to the thermal reservoir and directly proportional to the heat capacitance of the cell.
The MAC-E filter cut-off precision sets the energy range below the endpoint that is needed to maintain high
efficiency at the endpoint which in turn determines the total rate of $\beta$-decay electrons that register on the calorimeter.
The hit occupancy per calorimeter cell is set by the total incident flux divided by the granularity.  The full channel count of calorimeter cells
needed to cover the full tritium target exposure is in the upward range of $10^5$ channels, each with a bandwidth of
approximately 1~MHz.  A microwave-readout massive SQUID multiplexer system is being developed to 
handle the high channel capacity required for PTOLEMY.  An example of a MMSM layout is shown in
Figure~\ref{fig:SQUIDMultiplexer}.

\begin{figure}[h!]
\begin{center}
\includegraphics[width=0.75\textwidth]{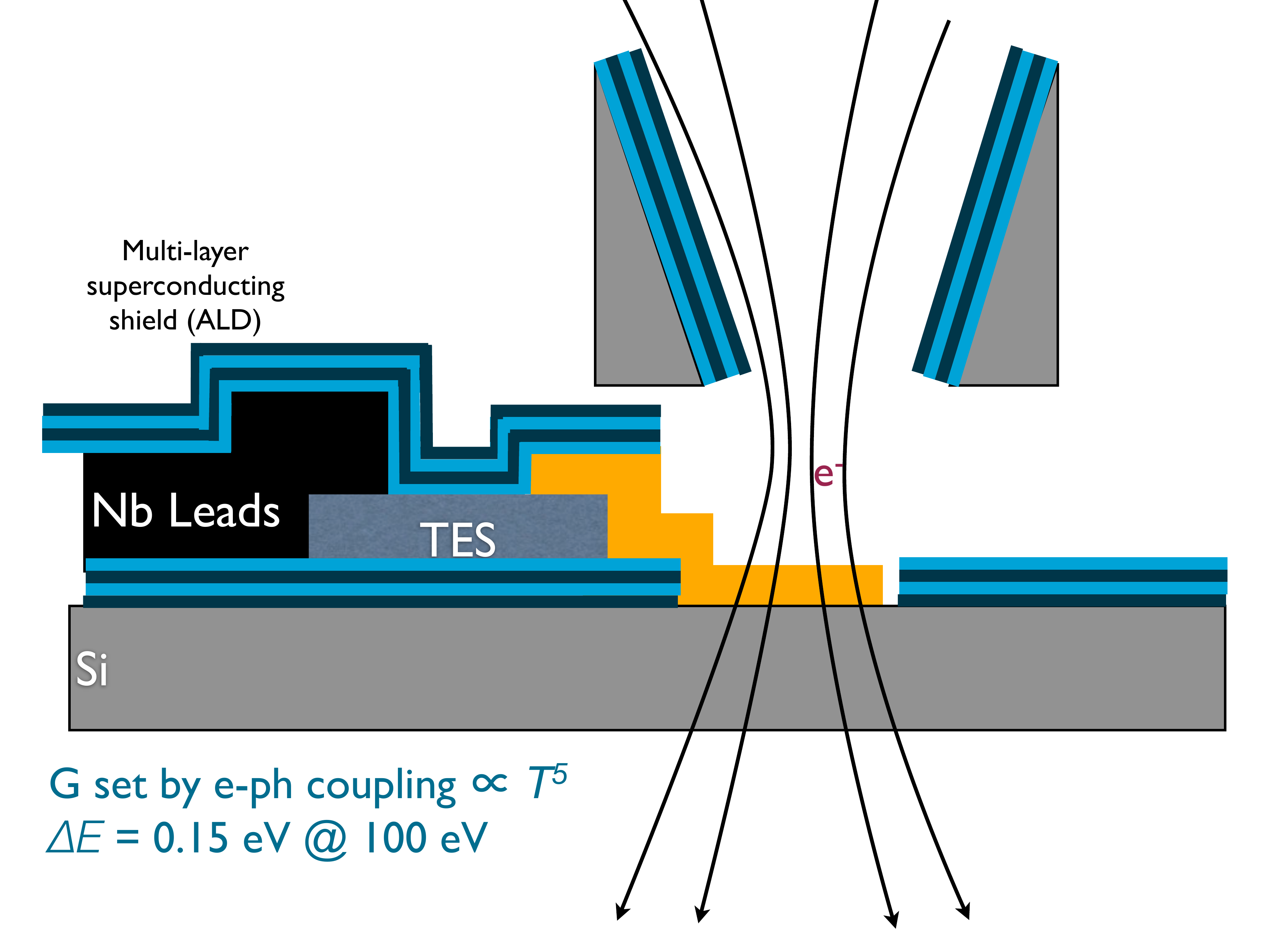}
\caption{Design concept for a high resolution, transition-edge sensor(TES) with normal regions 
to allow for electron transport on magnetic field lines that thread through the thermal mass
of the calorimeter.
\label{fig:TESlayout}}
\end{center}
\end{figure}

The magnetic field associated with operating transition-edge sensor(TES)-based microcalorimetry in conjunction
with a MAC-E filter is a major technical challenge.  TES readout systems are typically operated in low magnetic
field environments due to a downward shifting of the transition-edge temperature in high fields.  The design of the TES for PTOLEMY incorporates magnetic shielding for the TES and the SQUID readout system.  Magnetic field lines are focused into normal regions of the calorimeter.  The focusing is achieved by the magnetic shield mask that
screens magnetic fields by thin superconducting multilayers in a horn configuration.
The normal regions of the calorimeter are thermally coupled  to the TES.  Initial estimates for the energy resolution
are based on 40--400$\mu$m pixel size operating at 70~mK, although there are many advantages in trying to
achieve lower operating temperatures and increasing the pixel size. 
A preliminary design of the superconducting mask is in progress.  Previous work with interspersing normal and superconducting regions in the TES at
Argonne National Laboratory(ANL) indicate that a suitable mask can be developed for PTOLEMY.
A preliminary design is shown in Figure~\ref{fig:TESlayout}.

%% file: sections/TriggerDAQ/triggerdaq.tex
\vspace{-0.25 truein}
\section{Trigger and Data-Acquisition}

The raw rate of electron production from tritium $\beta$-decay for 100grams
of tritium is roughly 10$^{16}$ electrons/second.  The fraction of $\beta$-decays
within 100~eV of the endpoint is approximately $2 \times 10^{-7}$.  This fraction scales as
$m^3_\nu$, the neutrino mass, and therefore within 0.1~eV of the endpoint, the event rate is 
$2 \times 10^{-16}$ of the raw rate, or approximately 2~Hz in the signal region.
With $10^5$ readout channels, the average rate per channel is 10--20~kHz.
The degree to which the distance to the endpoint can be reduce depends on the
precision of the MAC-E filter cut-off, with a target precision of $10^{-2}$--$10^{-3}$.
Therefore, bringing down the $\beta$-decay rate impinging on the calorimeter to
1~kHz per channel for 10$^5$ channels is possible within the range of design parameters 
for the MAC-E filter.

With a microwave-readout massive SQUID multiplexer(MMSM) capable of combining
10$^3$ channel of calorimeter readout, each is  approximately 1~MHz of bandwidth, 
the total calorimeter readout is expect to have approximately 100 MMSM readout channels,
each operating at 1--10~GHz.

The RF tracker will have the same occupancy as the calorimeter with 10$^5$ individual
waveguides each operating at 38--46~GHz with a bandwidth of roughly 100~MHz.
For approximately 1~$\mu$sec RF tracker transit times over a window of 10~$\mu$sec
of calorimeter-RF track coincidence, approximately 1000~samples of 10ns flash-ADC data
will need to be pipelined and fed through a digital trigger that evaluates the RF tracker
amplitude data, the calorimeter data, and data from the muon veto system.  The background
rejection algorithms will need to reject multi-hit backgrounds in regional regions, and suppress
timing windows where large background-induced high occupancy is observed.

%% file: sections/TOF/tof.tex
\vspace{-0.25 truein}
\section{Time-of-flight}

Time domain data are recorded by the RF tracker system and the calorimeter.  At this time there is no 
foreseen instrumentation of the graphene target for a possible timing coincidence through $^3$He recoil
detection or changes in the conductivity of the partially tritrated graphene substrate.  However, it is possible
to calibrate a common timing reference for the RF tracker system and calorimeter by periodically
stepping the extraction voltage on the tritium target and measuring the time delay on the corresponding
detection signals with electrons within 150eV below the tritium endpoint.  Time-of-flight (TOF) information constrain
the longitudinal and transverse velocity of the electron and therefore the production angle at the target.
This velocity information is largely independent of the amplitude measurement from the calorimeter.
Non-electron background rejection are strongly suppressed by requiring consistency in these measurements
within their experimental resolutions.

%% file: sections/MuonVeto/muonveto.tex
\vspace{-0.25 truein}
\section{Muon veto}

The segmented magnetic duct system with closed loop flux tubes provides a configuration for
PTOLEMY where separate tritium target containers can be located at varying distances to the
the opening of the MAC-E filter.  A modified, large diameter UC-609 container would meet the 
storage requirements for the tritium target and provide a compact volume that can be fully
enclosed in a muon veto system.  The magnetic duct and flux return can be fed horizontally
with a vertically oriented tritium target to minimize the cosmic ray flux normal to the target.

%% file: sections/Calibration/calibration.tex
\vspace{-0.25 truein}
\section{e-Gun Calibration}

PTOLEMY is partitioned into thousands of simultaneous, non-overlapping tritium endpoint measurements.
To combine these data, the relative energy scale and resolution of each individual endpoint measurement
must be calibrated in situ throughout the data collection period.  To maintain this calibration an electron gun
feds through a ferrule from the rear of each tritium target plate and propagates through the full PTOLEMY 
spectrometer.  A low-light-level, pulsed photocathode feeds the electron gun and only pulses liberating
single electrons are used for calibration.  The shot-to-shot energy spread is 0.1eV and the absolute energy
scale is maintained to within a fraction of the resolution.  A steering magnet is used to deflect transversely
the electron beam to scan the full range of emission angles from the tritium target.

%% file: sections/Vacuum/vacuum.tex
\vspace{-0.25 truein}
\section{Vacuum system}

The original target vacuum for the small scale PTOLEMY prototype was set at a modest 10$^{-5}$ torr range.  In order to fabricate and install the prototype system as economically as possible, a portable pumpcart from the vacuum prep lab at PPPL was used.  This pumpcart system is comprised of a turbomolecular pump (TMP), backing mechanical pump, 1000 torr capacitance manometer gauge, ion gauge and appropriate valving.  A quadrupole mass spectrometer RGA was also installed on the pumping system.  The system also is interlocked to close all valves and safe the system, including the PTOLEMY vacuum chamber in the event of a loss of power.  To eliminate the effects of the magnetic fields on the TMP and gauging, the pumping system is located approximately 6' from the vacuum chamber.  For ease of decoupling the vacuum system, it is connected to the vacuum chamber using a flexible bellows line.  Since the ion gauge or RGA cannot be directly on the vacuum chamber, a test was performed to measure the pressure gradient between the chamber and the pumping system.  As a result, it was determined that the chamber is approximately at a pressure 10 times higher than the ion gauge reading at the pumpcart.  The vacuum chamber was initially pumped down and a thorough leak check of the system was performed.  The system was disassembled and the leaks repaired and the entire system pumped down for a second time and the system pumped down to approximately $1\times 10^{-6}$~torr on the ion gauge ($\sim 1 \times 10^{-5}$~torr at the chamber).  At this pressure less than 10\% of the beta particles would be scattered over a 3 meter trajectory during the initial experiments.

A 10'' flange was installed on the main chamber (left hand side of fig.~\ref{fig:PTOLEMYphoto}) and the pumping line could easily be reconfigured for an 8'' pumpduct using a short flex line and stainless steel tubing.   For follow-up experiments, with the purchase of a larger TMP ($\sim$500 liter/s), the pumping speed could be increased by a factor of greater than 50 with a corresponding decrease in chamber pressure.  At these chamber pressures, the
transmission probability would be greater then 99\% for a $\beta$ particle to travel a path length of 3~meters.

%% file: sections/Cooling/cooling.tex
\vspace{-0.25 truein}
\section{Cooling systems}

The superconducting coils for the MAC-E filter are maintained at cryogenic temperatures with a cryostat.  These standard commercial cryostat systems consist of a liquid helium container surrounded by a liquid nitrogen container and thermal radiation insulation.  The entire cryostat is then maintained under vacuum to greatly reduce the heat transfer from thermal conductivity between the components of the cryostat.

For the prototype PTOLEMY experiments, the detector will be held at less than 1 K in a vacuum chamber at high vacuum.  To maintain the detector at this temperature, it will be installed on a cold finger which is mounted on a dilution refrigerator (DR).  Because this experiment requires the detector to be at sub K temperatures for the detection of beta particles from a remote location, in contrast to the experimental sample being located in the DR chamber itself, modifications are required from a typical DR and cryostat.  In standard DR operation, the DR is immersed in a large liquid helium container, but this is not possible for this experiment.  The DR will need to be configured with a compact, in-vacuum, cryostat with a small opening through it.  With this configuration, the mixing chamber and cold finger chamber will be kept at cryogenic temperatures and the cryostat will further shield the detector from nearly all but straight on line-of-sight radiation heat transfer.  While this represents a modification to a standard DR, similar types of DRs have been commercially manufactured.

The experiment also requires the source and the RF tracker components be maintained at cryogenic temperatures.  This will be done using the cryostat systems of two additional superconducting magnets.  The central bores of the superconducting magnet systems will be removed and the cryostat vacuum chambers will be configured as part of the experimental vacuum chamber.  With this design modification, both the source and the RF tracker will be supported and cooled to cryogenic temperatures with copper brackets that are in good thermal contact with the liquid helium reservoirs of the cryostats.

%% file: sections/Monitoring/monitoring.tex
\vspace{-0.25 truein}
\section{HV system and Conditions monitoring}

\begin{figure}[h]
\begin{center}
\includegraphics[width=0.65\textwidth]{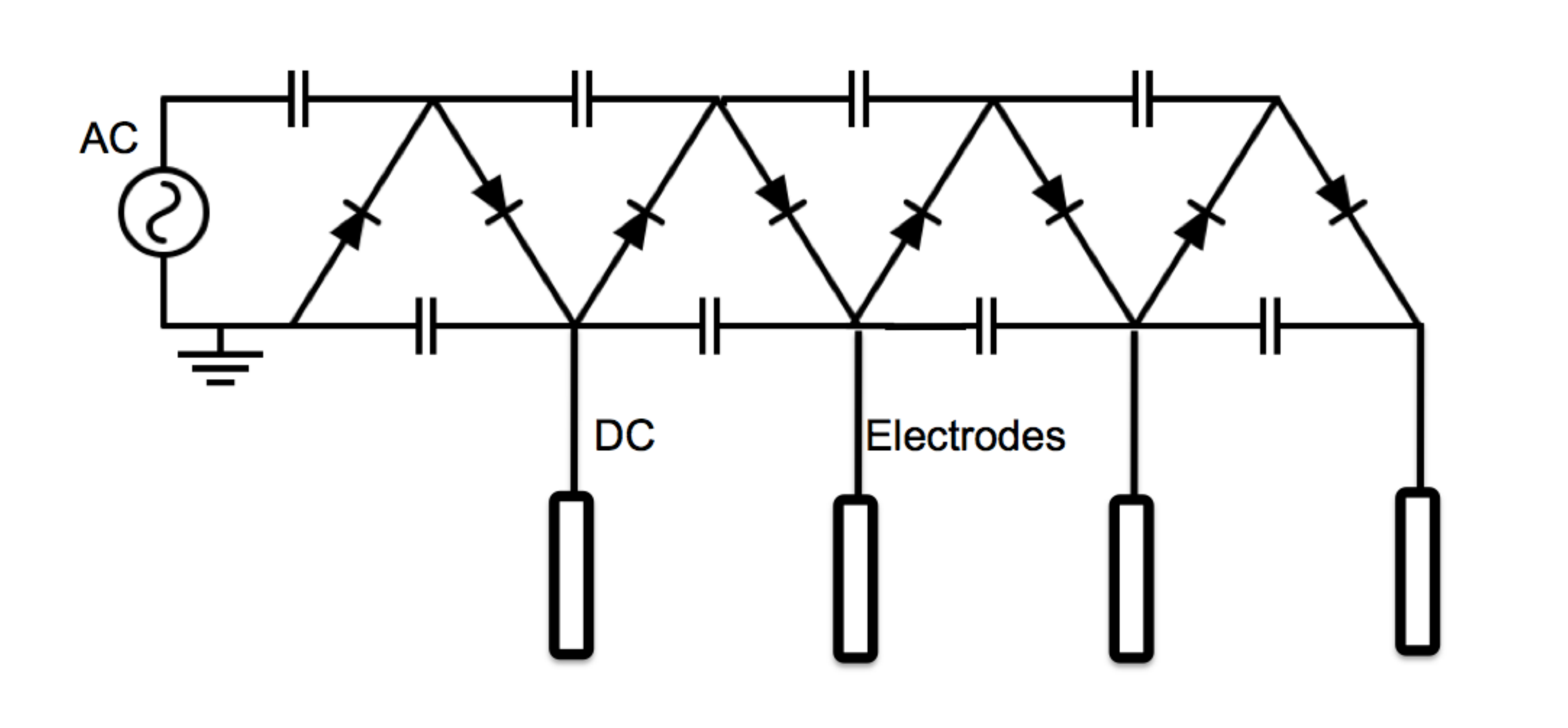}
\caption{Schematic of the Cockcroft-Walton (CW) circuit and electrodes for the HV system.
\label{hvcircuit}
}
\end{center}
\end{figure}

The stability of the HV system directly contributes to the absolute energy resolution of the electron measurement.  The intrinsic voltage stability over the trajectory of the electron from target to calorimeter needs to be better than 0.1~V.  A precision monitoring system with sensitivity of 1~meV can correct for slow leakage current drift of the voltage potentials.
For this purpose, we chose to use a Cockcroft-Walton (CW) circuit as a drift voltage generator that will be operated without any current load. 
In this circuit, an AC voltage is transformed to DC by means of a series of rectifying cells mounted in cascade, as one can see in the Figure~\ref{hvcircuit}.

Thus, the voltage stability is assured by blocking the charge on the capacitors and by the reverse current of the diodes keeping as low as possible.
Given that the overall voltage is not particularly high in this application (~18 kV) the CW can be designed with as many cells as the number of electrodes installed to make the electric field.  Then the connection of each electrode to the output of each cell will provide the desired voltages to bias the electrodes.
For the installation of the circuit, two possible configurations are foreseen.  One option, probably not the most favorable, is to install the circuit inside the drift volume and use the electrodes themselves to fasten the circuit. In this case, only the AC (500 VAC) pumping up the DC voltage must be introduced in the drift volume. 
In the other configuration each electrode is connected to a vacuum tested HV feed-through and which is connected on the air-side to one cell of the CW. The advantages of this configuration are that the full CW can be immersed in an insulation resin or silicon oil to prevent possible discharges or leakage current, both of which lower the performances of the circuit.  Further any possible intervention on the system would be easier and would not require accessing the inner volume of the detector.

As already pointed out, the monitoring of the voltage is of crucial importance.  Furthermore, any possible resistive load would affect the voltage measurement value, so a non-disruptive measurement method is desirable.  To meet the requirements, we can adopt a Field-Mill (FM) device that is already used in other detectors with similar requirement. The FM will be installed outside of the drift volume in any configuration of the HV circuit.

\begin{figure}[h]
\begin{center}
\includegraphics[width=0.65\textwidth]{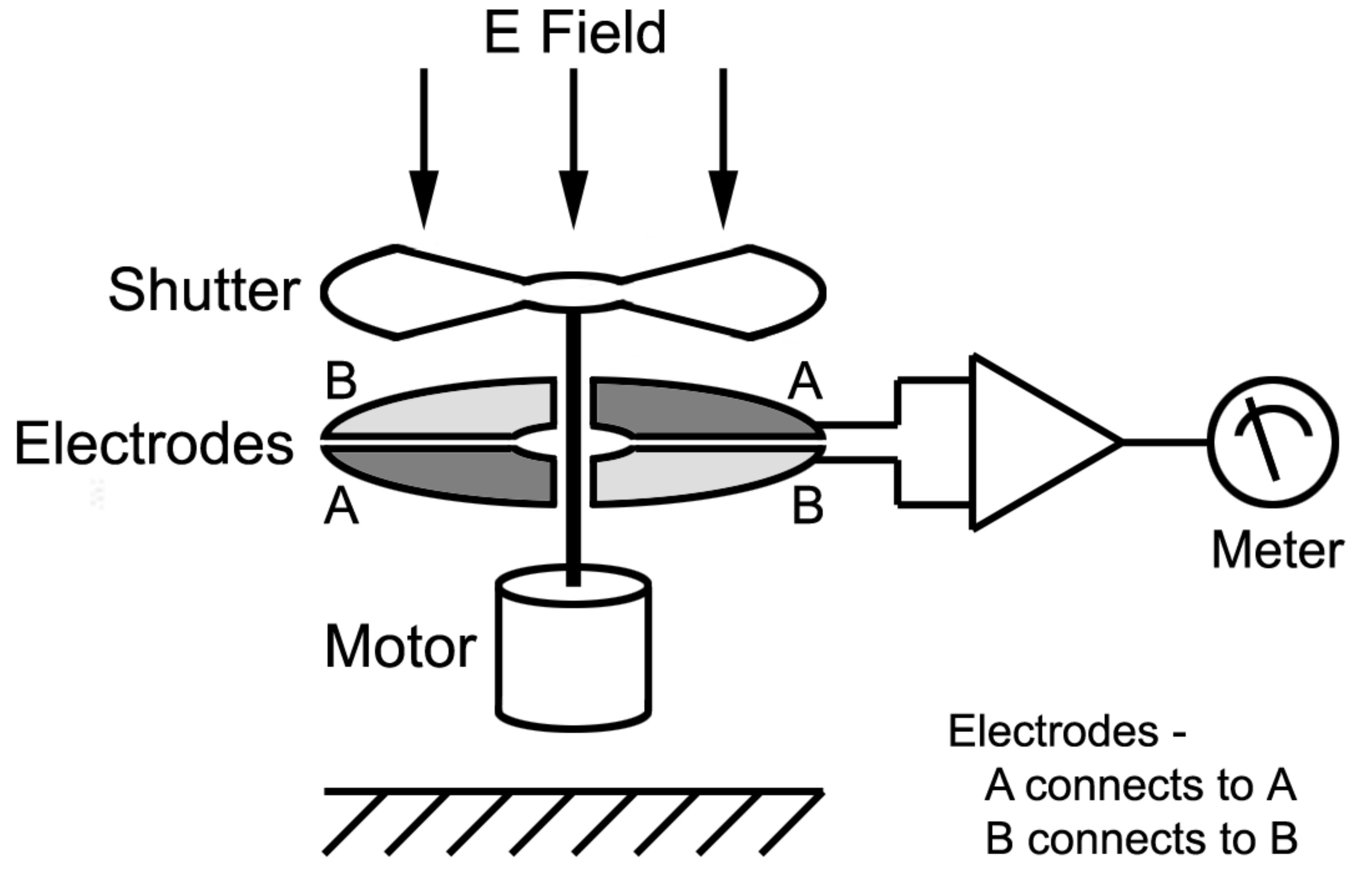}
\caption{Schematic representation of the Field-Mill voltage monitoring system.
\label{fm}
}
\end{center}
\end{figure}

In a FM, the voltage monitoring is obtained thanks to stators (Figure~\ref{fm} electrodes A and B) connected to a virtual ground of a charge amplifier and a rotor (Figure~\ref{fm} shutter), the latter activated by a motor. The whole setup is immersed in the E field generated by the voltage of the electrodes we aim at monitoring. When the rotor of Figure~\ref{fm} moves in front of the stators a pulse voltage will be generated, proportional to the charge induced on the stators by the E field. 
Once properly calibrated, the pulse voltage will provide the measurements of the electrodes voltage with the desired uncertainty (mV) and even better. 
With this method, a continuous monitoring of the voltage is also naturally realized.

The conditions data will be used to detect voltage jumps and discharge instabilities that affect the data quality.

%% file: sections/Support/support.tex
\vspace{-0.25 truein}
\section{Assembly, Support, and Maintenance}

The PTOLEMY proposal harnesses the unprecedented capabilities of a facility previously used to host the worldÕs most powerful fusion reactor, known as the TFTR.  The test cell facility was extensively developed for radiation and safety controls with extensive loading and assembly infrastructure.  Without the substantial capabilities of the existing infrastructure made available to this project, the vision of the PTOLEMY development would not be feasible.  The primary components of the PTOLEMY project, the Tritium source and vacuum chamber leverage previous infrastructure made available for this project.

%% file: sections/Safety/safety.tex
\vspace{-0.25 truein}
\section{Safety systems}

The PTOLEMY experiment uses cryogens, liquid nitrogen and liquid helium, and high static electromagnetic fields, which create hazards that could have serious consequences if not properly controlled. Control measures have been implemented for each identified hazard.

The use of cryogens creates three hazards: over pressurization of containers, the extreme cold, and the sudden boil-off of cryogenic liquids, called a quench. To prevent over pressurization, cryogens are contained in dewars with pressure-relief devices. Employees could come into contact with liquid helium or liquid nitrogen when filling the magnets. Only employees trained in the Safe Handling of Compressed Gases \& Cryogenic Liquids are authorized to do this task; face shields, goggles, and insulated gloves are worn when handling cryogenic fluids. Lastly, since liquid nitrogen and liquid helium expand by a factor of approximately 700 to 1, a sudden boil-off of these liquids has the potential to create an oxygen deficiency. Local exhaust hoses (``elephant trunks'') have been placed at the point of boil-off on both magnets. The trunks run continuously and their flow rates are checked annually. In addition, an oxygen monitor has been placed between the two magnets, alarming if an oxygen deficiency occurs. Workers have been trained to evacuate and contact Emergency Services should a low-oxygen alarm sound. The monitor is calibrated quarterly per the manufacturer instructions. 

According to the American Conference of Governmental Industrial Hygienists (ACGIH), Threshold Limit Values (TLVs), an employee can work in static magnetic fields with flux densities above 2 Tesla (T) but below 8 T if the employee has special training and works in a controlled workplace environment. The magnets used in this experiment are capable of having magnetic fields above 2 T but cannot reach 8 T. All employees working on the PTOLEMY experiment have completed High Static Magnetic training, which addresses both the biological and mechanical effects of high static magnetic fields. 

A number of control measures have been implemented to ensure that the experiment is operated in a controlled environment. The experiment is located inside a locked enclosure and only employees with the High Static Magnetic Field training can gain access. A sign is posted next to the enclosure door, reminding individuals that ferromagnetic objects are not allowed inside. Inside the enclosure, a red line is painted at 50 Gauss. Escorted, untrained individuals are not permitted to cross this line. 

ACGIH Ceiling Value for medical device wearers is 0.5 mT or 5 Gauss. The front of the enclosure lies at the 5 Gauss limit. A sign reading, ``DANGER Magnetic Fields Persons with Medical Implants Do Not Enter'' is posted on the enclosure door, warning individuals with medical implants not to enter.  Danger tape with the same sign hangs across the rest of the 5 Gauss line perimeter (where the line is not protected by the enclosure), acting as a barricade. 

A Job Hazard Analysis (JHA) was completed to encompass all of these hazards and is kept inside the enclosure. The JHA is updated for each phase of the experiment. All personnel working on the experiment have participated in a pre-job brief, detailing the contents of the JHA and the control measures listed above.

%% file: sections/ScienceEd/science.tex
\vspace{-0.25 truein}
\section{Education, Outreach, and Knowledge Transfer}

For more than 20 years, science education programs organized by members of this proposal have trained the next generation of scientists and engineers through undergraduate research internships.   The successes of previous science education programs run by members of this proposal provide a successful model for how to build a new program focused on the unique research opportunities related to the development of a relic neutrino detection
experiment.
We expect that the PTOLEMY facility will train 10 - 15 students each year.  Students will be recruited from universities throughout the country and spend nine weeks in the summer at the research development facility in Princeton, receiving a stipend, and housing/travel costs. 
Recent studies, documented below, have found that a positive science education internship that replicates an authentic research experiences can significantly reinforce a studentÕs desire to continue their studies in graduate school.   Close collaboration with a mentor is crucial but we also require each student at the end of the summer to write a report and make an oral presentation.   In addition, our previous science education programs go further than most in replicating this authentic experience by having students present their work at a professional conference.  In our previous programs, students present their work at an American Physical Society conference and the program covers the travel costs.  A ten-year longitudinal study of career choices after existing participants completed an internship is documented in Life Sciences Education Vol. 6, 297Ð306, Winter 2007, Undergraduate Research Experiences Support Science Career Decisions and Active Learning by David Lopatto.  In what follows, we highlight the successes of previous programs in comparison with national averages and emphasize that these improvements indicate a strong positive aspect of these programs in agreement with the general conclusions of the Lopatto article.

Students completing an internship at our previous science education programs go to graduate school at a frequency of nearly a factor of two when compared to national averages.  In addition, nearly twice as many women compared to the national average that completed a science education internship organized by members of this proposal go on to graduate school.
Comparing the statistics from the American Institute of Physics with those of students completing previous science education programs organized by members of this proposal:
\begin{itemize}
\item[$\bullet$] 64\% of all program participants go on to a graduate program in physics, math, or engineering as compared to 1987-2003 national average of 35\%.
\item[$\bullet$] From 2002 Ð 2005, 30\% of all program participants that entered a physics graduate program were female as compared to the 2004 national average of 16\%.
\end{itemize}

Members of this proposal have been consistently recognized as "Outstanding Mentors," receiving 9 awards for previous science education programs.  A total of 22 undergraduate students from previous programs have won "best poster" awards at the APS DPP conference (out of 40 given out) in the past 10 years.

A partial list of schools of past participants shows broad geographic diversity:

Univ. of Washington, UCLA, Union College, Rice Univ., Howard Univ., Univ. of Minnesota, Cornell Univ., Univ. of Wisc-Madison, Clemson Univ., St. Peters College, Seton Hall Univ., Drew Univ., Univ. of Chicago, SE Missouri State, Princeton Univ., Boston Univ., Ashland Univ., Whitworth Univ., The College of NJ, MIT, Mesa State College, Univ. of San Diego, Vanderbilt Univ., Purdue Univ., RPI, Montana State Univ., Utah State Univ., Embry-Riddle Aeronautical Univ., Ohio State Univ., UC-Berkeley, Univ. of Calif. Ð Irvine, Franklin \& Marshall, Butler College, Univ. of Penn, Univ. of St. Thomas, Univ. of SoCal, Arizona State Univ., Utah State Univ., Univ. of Colorado-Boulder, Wheaton College, Kansas State, Washington \& Lee Univ., Hendrix College, Pacific Lutheran Univ., Brown University, Vassar College, Florida State Univ., Texas Christian Univ., Delaware State Univ., Colorado School of Mines, Truman State Univ., Ohio Wesleyan Univ., Saint Louis Univ., Columbia Univ., Walla Walla College, Univ. of Tulsa, Colby College, Univ. of Rochester, Univ. of Missouri-Rolla, Stanford University, Univ. of Rochester, GA Inst. of Tech.

\subsection{Broadening Participation}

It is too often the case that science education internships terminate the relationship between the researchers and the students when the program ends.  That is not the case with the programs run by members of this proposal.  In fact, building a relationship between an individual researcher and a student is more valuable than the specific science education training that they receive during their first encounter in the program.  The impressions of community and scientific integrity are strong values that are part of the research environment and ones that can broaden the outlook of new students into pursuing an interest in science and engineering.  The programs members stay at open arms to future, current and previous program participants to nurture their development in the sciences and to build a more diverse and far-reaching future for the scientific fields.  There are a number of concrete examples where previous participants have returned and engaged in long-term research programs with facilities involved in the science education programs.
The PTOLEMY science education program will benefit from a new collaboration with Howard University building on a program that brings undergraduate students and faculty from minority serving institutions to the laboratory to work on research projects of mutual interest.  The current project in which students and faculty of Howard University are researching, stemming from a previous program, is centered upon the design and testing of new lubricants in ultra high vacuum (UHV) environments and has direct application to technologies that will become available in the PTOLEMY project.  It is the commitment of the PTOLEMY researchers to continue to engage and involve broader participation and to establish long-term research relationships with underrepresented universities so that their universities can grow and diversify their opportunities for advancement in science for their own students.  We anticipate 2-3 undergraduates and 1-2 faculty members from Howard or other Minority Serving institutions to participate in these activities each year.
In addition, Princeton University has also been an employer for students from Drexel UniversityÕs Undergraduate Cooperative Education Program. This program enables undergraduate students to balance classroom theory with practical, hands-on experience prior to graduation through working at the laboratory for six months of the year. The students in this program are either in their sophomore, pre-junior, or junior year in their undergraduate education, generally studying engineering or computer science.   These students will be incorporated into the PTOLEMY project at all levels at a level deeper than a typical and shorter summer internship.  We anticipate 2-3 Drexel students participating in these activities each year.

%% file: sections/Prototype/prototype.tex
\vspace{-0.25 truein}
\section{PTOLEMY prototype}

\begin{figure}[h!]
\begin{center}
\includegraphics[width=\textwidth]{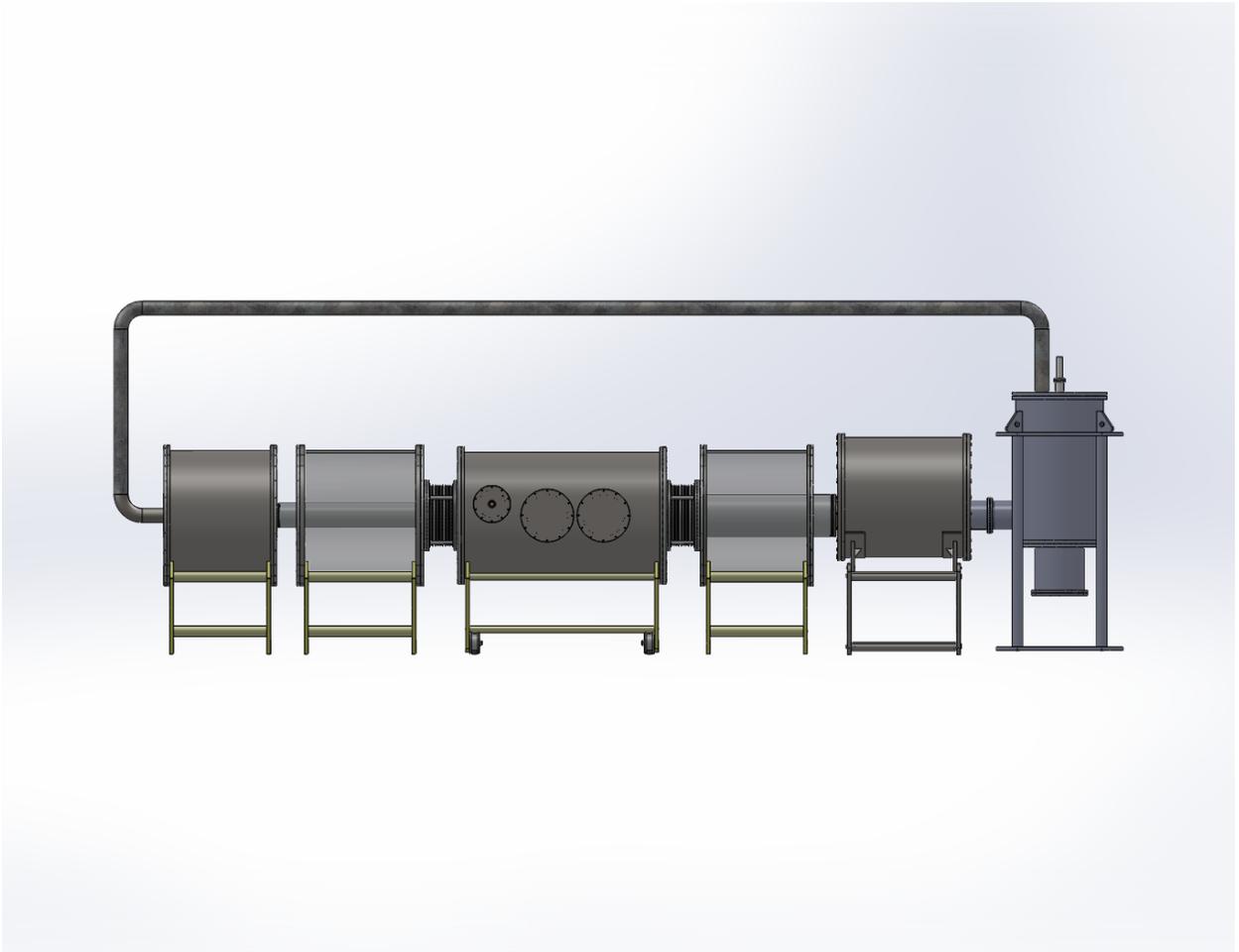}
\caption{The PTOLEMY prototype design starts with a 4K surface-deposition tritium source in the left
LHe-cooled vacuum chamber, accelerates into a MAC-E filter with $10^{-2}$ cut-off precision, accelerates 
electrons above the endpoint and down to 180eV below the endpoint into a uniform field solenoid where the RF signal from the cyclotron motion of individual electrons in a 1.9T magnetic field provide a tracking detector measurement above a minimum transverse momentum, the 4K cooled RF waveguides extend into the adjacent Ion Cyclotron Resonance magnet to provide a second tracking segment, then finally the electron is decelerated into a sub-keV energy range, low magnetic field region, and measured with a high resolution cryogenic calorimeter in a 100mK dilution refrigerator (DR) and put in time-of-flight coincidence with the RF tracker.  A solid iron pipe returns
the magnetic field lines in a closed loop.
\label{fig:PTOLEMYupgrade}}
\end{center}
\end{figure}

\begin{figure}[h!]
\begin{center}
\includegraphics[width=0.6\textwidth]{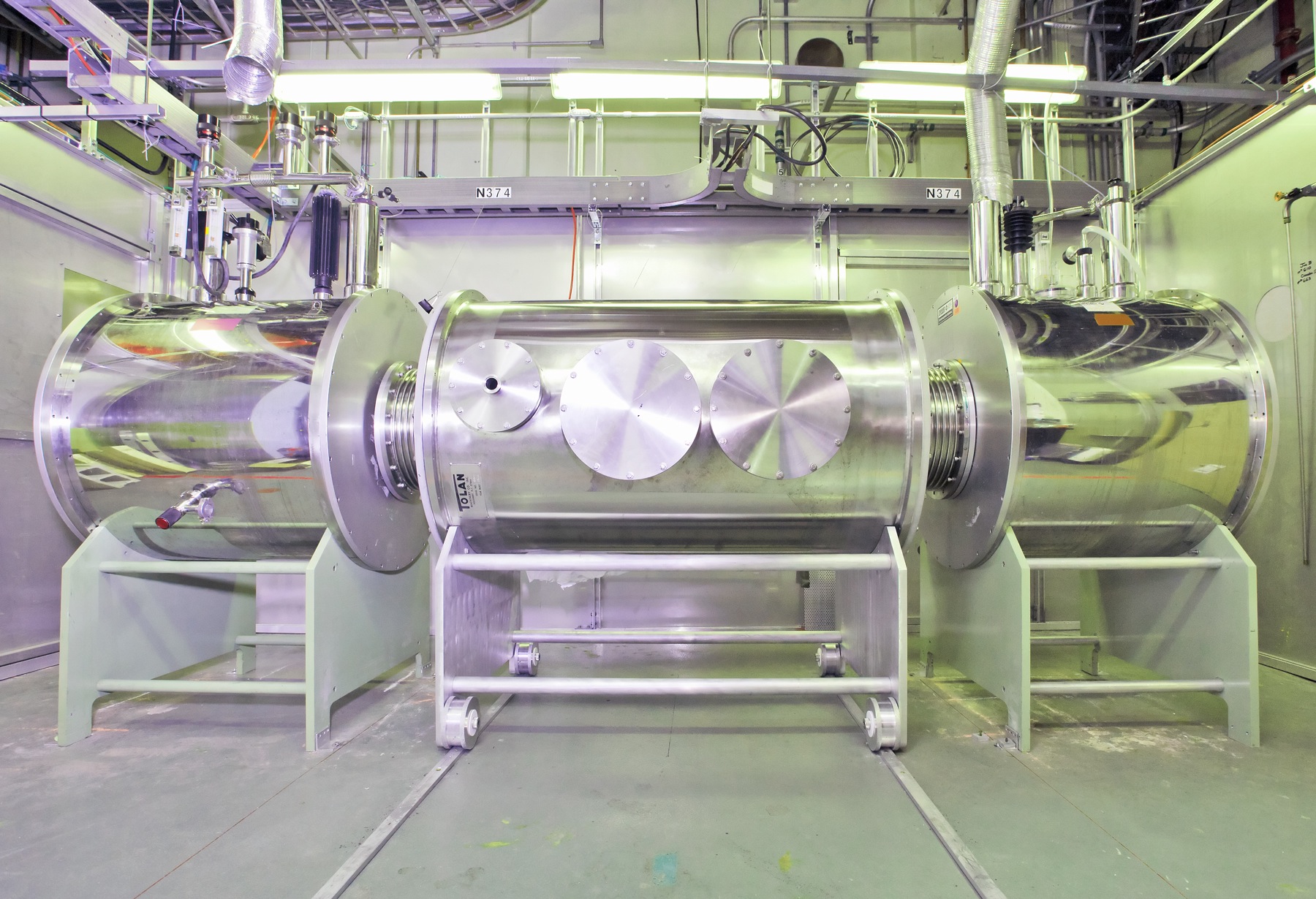}
\caption{The small-scale PTOLEMY prototype installed at the Princeton Plasma Physics Laboratory (February 2013).  Two horizontal bore NMR magnets are positioned on either side of a MAC-E filter vacuum tank.  The tritium target
plate is placed in the left magnet in a 3.35T field, and the RF tracking system is placed in a high uniformity 1.9T field
in the bore of the right magnet with a windowless APD detector and in-vacuum readout electronics.
\label{fig:PTOLEMYphoto2}}
\end{center}
\end{figure}

The PTOLEMY prototype is a compact MAC-E filter with a tritium target on the input and an RF tracker and cryogenic microcalorimeter on the output.  The prototype will operate a complete high precision test string of the PTOLEMY
technologies to provide the first data on the intrinsic precision on the energy of endpoint electrons from atomically bound tritium on graphene, the single electron RF signal-to-noise and timing precision in a tracker configuration, and the cryogenic calorimeter energy measurement resolution for sub-keV electrons in a magnetic field environment.  
A cryogenic calorimeter with sub-eV energy measurement resolution and sub-eV energy smearing of the tritium 
target are the two most fundamental new capabilities of the PTOLEMY project.
The planned upgrade to the existing setup is shown in Figure~\ref{fig:PTOLEMYupgrade} where at the moment
the central vacuum chamber has been installed and pumped down and the two adjacent 
superconducting magnets are installed and were energized, as shown in Figure~\ref{fig:PTOLEMYphoto2}.
The prototype system will collect the tritium endpoint spectrum from a number of potential tritium target designs 
with graphene and provide physics-quality data with unique search sensitivity for 
eV to keV mass sterile neutrinos that may have been produced in the early universe.
The final precision on the electron endpoint energy and component-wise evaluation of the background
levels in the prototype system will be used to model and extrapolate the prototype design
to a system with 100gram-year exposure capability.  The precision on the tritium target and the
required background levels will provide an accurate assessment on the feasibility of a full-scale experiment with the 
goal of cosmic relic neutrino detection.

The superconducting magnets and MAC-E magnetic geometry have been validated with a magnetic
field map with both magnets energized.   The mid-plane field is approximately 300~Gauss, approximately
1\% of the high field magnet.  Currently, the high field magnet is energized and has been 
maintained in stable operation for over one year.  Flash-ADC readout electronics and a windowless APD 
electron detection system for background measurements have been operated in the high field magnet for initial testing of the HV system
voltage offset on the endpoint of a $^{14}$C source.  The 1.9T large bore magnet is powered down and will be 
re-energized with the MRI shim coils adjusted to provide a uniform magnet field to 1 part in $10^{5}$ over a 10~cm
length along the bore axis.
In this region, the Q-band magic tee waveguide will be installed and cooled down to 4.2K using the
inner cryo-wall of the superconducting solenoid integrated into the MAC-E filter vacuum.  
A longer, higher uniformity magnetic field can be produced with a superconducting Ion Cyclotron Reasonance(ICR) magnet.  The ICR will also be incorporated into the PTOLEMY prototype in tandem with the MRI magnet
for more detailed studies of the RF signal-to-noise.  The electrons after having passed through the MRI and
ICR magnets will be guided vertically through a magnetic-field tolerance dillution refrigerator(DR) with
a cryogenic microcalorimeter.  Finally, the flux for the magnetic field will be returned through an iron yoke
to the tritium target.

\begin{figure}[h!]
\begin{center}
\includegraphics[width=0.5\textwidth]{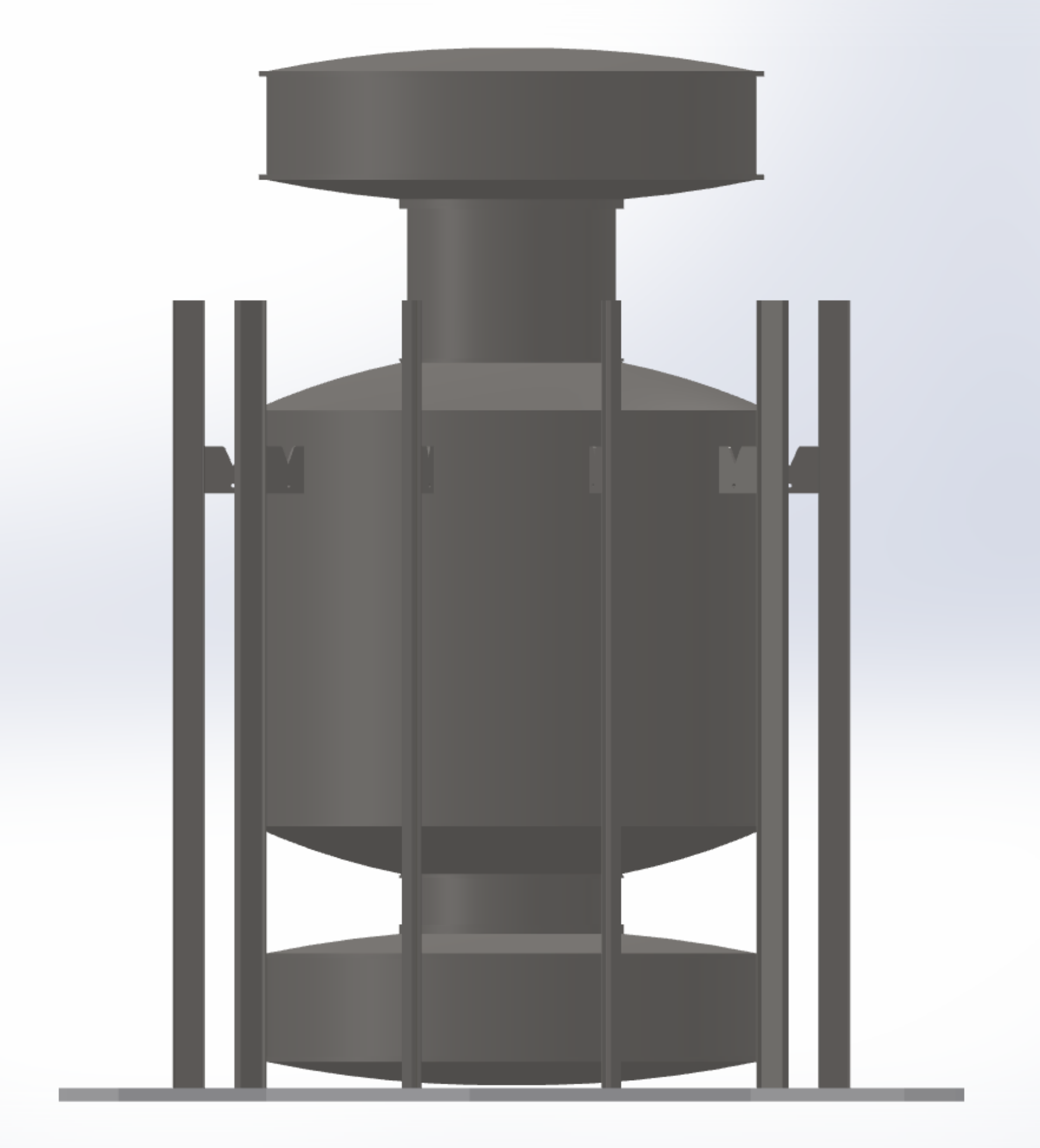}
\includegraphics[width=0.35\textwidth, height=9cm]{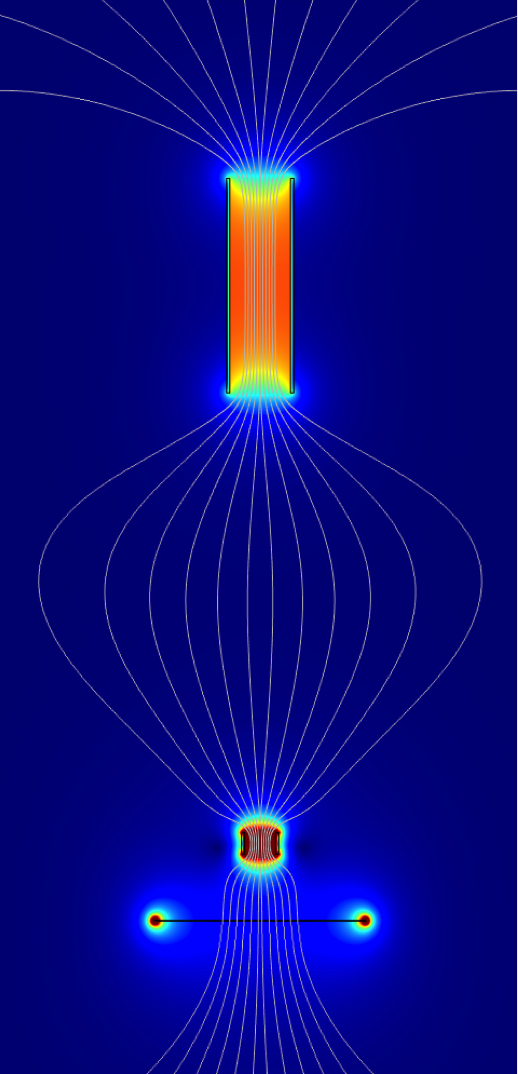}
\caption{The 100gram PTOLEMY single-chamber structure is shown on the left and the corresponding magnetic
geometry is shown on the right.
\label{fig:PTOLEMYfullscale}}
\end{center}
\end{figure}

\begin{figure}[h!]
\begin{center}
\includegraphics[width=0.85\textwidth]{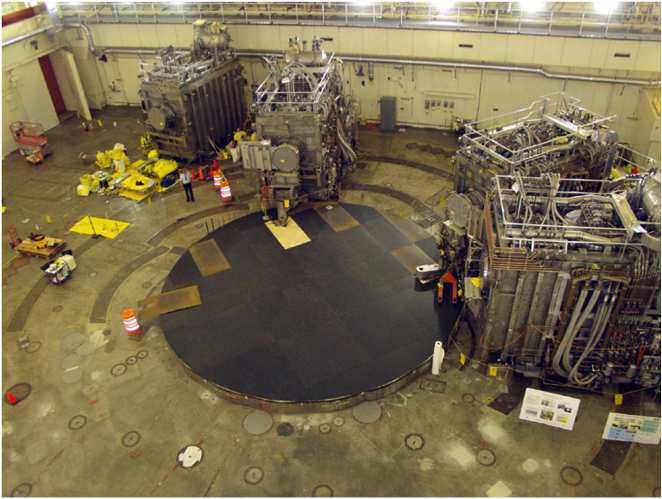}
\caption{The TFTR test cell with the round pedestal above the test cell basement shown at the center.
\label{fig:testcell}}
\end{center}
\end{figure}

In parallel with the precision measurement test string, a high resolution, high capacity
tritium target will be built in the test cell basement.  A single-chamber, large scale concept for the 
PTOLEMY is shown in Figure~\ref{fig:PTOLEMYfullscale} with the mechanical structure and magnetic 
geometry.  There are several constraints with a single-chamber design that increase the vertical height
of the experiment.  The dilution refrigeration for the cryogenic calorimetry
requires gravity and necessarily is located at the top of the experiment.
Therefore, the lowest part of the structure is the tritium target.  This target can be located in the test cell
basement with the first magnet of the MAC-E filter placed in the portal between the test cell basement and 
the test cell, shown in Figure~\ref{fig:testcell}.

The segmented magnetic duct system has many advantages for PTOLEMY relative to a
single-chamber design.  With a magnetic duct geometry, the test cell space can
accommodate a larger volume MAC-E filter.  A dome geometry for the MAC-E filter has been shown 
to provide a uniform cut-off for multiple magnetic ducts.  With multiple magnetic ducts,  
the cryogenic calorimetry can be separated into individual DR units and placed at floor level.
The calorimeters are connected to the MAC-E filter via the magnetic ducts
with a return flux feeding the back of the corresponding tritium target containers.  The long magnetic ducts 
between the MAC-E filter and calorimeter are instrumented with the RF tracker system and
therefore provide long transit times and thereby decrease the bandwidth and increase the
signal-to-noise and timing precision.  Similarly, the tritium containers are separate and individually 
surrounded by muon veto systems below ground with increased concrete overburden to reduce backgrounds entering the tritium target region.

The future direction of the PTOLEMY design for 100gram-year exposure capacity will profit
from the upgrade prototype system with its program of technology validation and background 
assessment and from continued design studies that evolve with the feedback from the prototype studies,
full simulation and analysis, and related progress in the instrumentation R\&D.